\documentclass[11pt,a4paper]{article}
\pdfoutput=1

\usepackage{jheppub}
\usepackage[T1]{fontenc}
\usepackage{extarrows}

\title{Linearly resummed hydrodynamics in a weakly curved spacetime}
\author{Yanyan Bu}
\author{and Michael Lublinsky}
\affiliation{Department of Physics, Ben-Gurion University of the Negev, \\
Beer-Sheva 84105, Israel}

\emailAdd{yybu@post.bgu.ac.il}
\emailAdd{lublinm@bgu.ac.il}

\abstract{We extend our study of all-order linearly resummed hydrodynamics in a flat space~\cite{1406.7222,1409.3095} to fluids in weakly curved spaces. The underlying microscopic theory is a finite temperature $\mathcal{N}=4$ super-Yang-Mills theory at strong coupling. The AdS/CFT correspondence relates black brane solutions of the Einstein gravity in asymptotically \emph{locally} $\textrm{AdS}_5$ geometry to relativistic conformal fluids in a weakly curved 4D background. To linear order in the amplitude of hydrodynamic variables and metric perturbations, the fluid's energy-momentum tensor is computed with derivatives of both the fluid velocity and background metric resummed to all orders. We extensively discuss the meaning of all order hydrodynamics by expressing it in terms of the memory function formalism, which is also suitable for practical simulations. In addition to two viscosity functions discussed at length in refs.~\cite{1406.7222,1409.3095}, we find four curvature induced structures coupled to the fluid via new transport coefficient functions. In ref.~\cite{0905.4069}, the latter were referred to as gravitational susceptibilities of the fluid. We analytically compute these coefficients in the hydrodynamic limit, and then numerically up to large values of momenta.}


\arxivnumber{TBA}
\begin{document}
\maketitle

\flushbottom
\section{Introduction}\label{section1}
Fluid dynamics~\cite{fluid1,km,fluid2} is an effective long-distance description valid for most classical or quantum many-body systems at nonzero temperature. It is defined in terms of thermodynamical variables.  Derivative expansion in fluid-dynamic variables such as velocity accounts for deviations from thermal equilibrium.  At each order, the derivative expansion is fixed by thermodynamic considerations and symmetries, up to a finite number of transport coefficients, such as viscosity and conductivity. The latter  are not calculable from hydrodynamics itself, but have to be determined from underlying microscopic theory or experimentally.

The stress-energy tensor $T_{\mu\nu}$ of a relativistic fluid is conveniently written as
\begin{equation}\label{st result1}
\begin{split}
T_{\mu\nu}=(\varepsilon+P)u_{\mu}u_{\nu}+Pg_{\mu\nu}+\Pi_{\langle\mu\nu\rangle},
\end{split}
\end{equation}
where $\varepsilon$, $u_{\mu}$ and $g_{\mu\nu}$ are the fluid energy density, four-velocity and metric tensor. The microscopic theory underlying the hydrodynamics is supposed to specify the fluid pressure $P$ via equation of state, $P=P(\varepsilon)$. The viscous effects are encoded in the dissipation tensor $\Pi_{\langle\mu\nu\rangle}$,
\begin{equation}
\Pi_{\langle\mu\nu\rangle}\equiv\frac{1}{2}\Delta_{\mu}^{\alpha}\Delta_{\nu}^{\beta}\left( \Pi_{\alpha\beta}+\Pi_{\beta\alpha}\right)-\frac{1}{3}\Delta_{\mu\nu}\Delta^{\alpha\beta} \Pi_{\alpha\beta},
\end{equation}
where, in the local rest frame, $\Delta_{\mu\nu}=g_{\mu\nu}+u_{\mu}u_{\nu}$ acts as a projector on the spatial subspace. For conformal fluids, the first order gradient expansion has one term only, the Navier-Stokes term
\begin{equation}\label{ns hydro}
\Pi_{\mu\nu}^{\textrm{NS}}=-2\eta_{0}\nabla_{\mu} u_{\nu},
\end{equation}
where $\eta_{0}$ is a shear viscosity. Eq.~(\ref{ns hydro}) relates the spatial components of the stress-energy tensor ($T_{ij}$) to time involving components $T_{0j}$.
A relation of this type is frequently referred to as a constitutive relation.
Energy/momentum conservation (conservation of $T_{\mu\nu}$) leads to fluid dynamical equations, which are normally solved as initial value problem, when initial profiles for temperature and velocity are specified. We shall generically refer to these equations as Navier-Stokes equations, even when we extend the discussion beyond the first gradient.

Although fluid dynamics has long history, theoretical foundations of relativistic viscous hydrodynamics are not yet fully established. The Navier-Stokes hydrodynamics introduced in~(\ref{ns hydro}) leads to violations of causality: the set of fluid dynamical equations makes it possible to propagate signals faster than light. To overcome this problem, simulations of relativistic hydrodynamics are usually based on phenomenological prescriptions of~\cite{Muller,Israel,IS1976,IS1979}, which admix viscous effects from second order derivatives, so to make the fluid dynamical equations causal. Refs.~\cite{Muller,Israel,IS1976,IS1979} introduced retardation effects for irreversible currents, which, via  equations of motion, become additional degrees of freedom. In other words, one needs to include higher order gradient terms in the derivative expansion in order  to obtain a causal formulation. In general,  causality is violated if the derivative expansion is truncated at any \emph{fixed} order. It is supposed to be restored when all order gradient terms are included, which we refer to as \emph{all order resummed} hydrodynamics. Causality usually also implies stability and even can be used to constrain possible values of higher order transport coefficients~\cite{Hiscock1983,Hiscock1985, Hiscock1987, 0807.3120,0907.3906,1102.4780}.

Discovery of AdS/CFT correspondence~\cite{hep-th/9711200,hep-th/9802109, hep-th/9802150}, and its generalizations, has established a novel approach for exploring a large class of quantum field theories at strong coupling. Within the holographic picture, hydrodynamic fluctuations can be regarded as gravitational fluctuations of black holes in  asymptotic AdS spacetime, and vice versa~\cite{hep-th/0205052,hep-th/0210220}. The connection between fluid dynamics and black holes in AdS gravity is usually referred to as fluid/gravity correspondence. One central result of this correspondence is the ratio between $\eta_{0}$ and the entropy density $s$~\cite{hep-th/0104066,hep-th/0205052,hep-th/0405231}
\begin{equation}\label{ratio}
\frac{\eta_{0}}{s}=\frac{1}{4\pi}.
\end{equation}
This ratio~(\ref{ratio}) is universal~\cite{hep-th/0311175,0808.1837, 0809.3808} for a large class of strongly interacting gauge theories for which holographic duals are governed by Einstein gravity in asymptotically AdS spacetime.

Remarkably, the fluid/gravity correspondence is not limited to linear response theory for small perturbations of the velocity field. \emph{Nonlinearly} generalized Navier-Stokes equations are completely encoded in the Einstein equations. Particularly, the formalism of~\cite{0712.2456} provides a systematic framework to construct  nonlinear fluid dynamics, order by order in the velocity derivative expansion, with the transport coefficients determined from the gravity side. The study of~\cite{0712.2456} was subsequently generalized to conformal fluids in higher dimensions~\cite{0806.4602}, weakly curved background manifolds~\cite{0809.4272}, and to forced fluids~\cite{0806.0006}. We refer the reader to~\cite{0704.0240,0905.4352,1107.5780} for comprehensive reviews of fluid/gravity correspondence.

In~\cite{1406.7222,1409.3095}, we built upon previous work of~\cite{0905.4069} and constructed a flat space \emph{all-order linearly resummed} relativistic conformal hydrodynamics using the fluid/gravity correspondence. We have collected all the derivative terms that are linear in the fluid dynamic variables, like $(\nabla\nabla\cdots\nabla u)_{\mu\nu}$; but we neglected all nonlinear structures, such as $(\nabla u)^2_{\mu\nu}$.  The velocity field $u_\mu$ has a big time component, $u_0\sim 1$, whereas the three velocity $\vec u$ is considered to be small. It is worth stressing  that the \emph{linear} derivative terms can be made dominant by restricting our study to small $\vec u$, hence our approximation is well under control. Ideologically similar to our study here, resummations of gradient terms for boost-invariant plasma were considered in~\cite{hep-th/0703243,0906.4423,1103.3452,1203.0755,1302.0697,1411.1969}, with the prime difference that non-linear terms were also accounted for in these papers.

In~\cite{1406.7222,1409.3095} relativistic hydrodynamics with all order derivatives resummed was found to have a rich structure, absent in a strict low frequency/momentum approximation. The fluid stress-energy tensor was expressed using the shear term (\ref{ns hydro}), with $\eta_0$ replaced by a viscosity function of space-time derivative operators, and a new viscous term, which emerged starting from the third order in the gradient expansion. In Fourier space, both coefficient functions accompanying these two terms become functions of frequency and spatial momentum. They were calculated in~\cite{1406.7222,1409.3095}. Those viscosity functions were found to vanish at very large momenta, which is a marking signature of causality restoration.

In the present work, we generalize computations of~\cite{1406.7222,1409.3095} by  including small metric perturbations in the boundary theory. For $\mathcal{N}=4$ super-Yang-Mills theory, we deduce the stress tensor for all-order linearized relativistic hydrodynamics in a weakly curved non-dynamical background spacetime. Following the method of~\cite{0712.2456, 0806.0006}, we construct asymptotically locally $\textrm{AdS}_5$ solutions to the Einstein gravity with a weakly curved boundary metric. However, our procedure is somewhat different from that of~\cite{0712.2456,0806.0006}, and it has been invented in order to include all order linear structures in a self-consistent manner. In particular, rather than considering order-by-order derivative expansion,
we will collect all the derivative terms in a unified way~\cite{1406.7222,1409.3095}. We first solve the dynamical components of the bulk Einstein equations, which is sufficient to compute an ``off-shell'' stress-energy tensor of the boundary fluid with the  transport coefficients  to be  determined completely. Putting ``on-shell'' thus obtained stress-energy tensor is equivalent to constraints of the Einstein equations.

The boundary metric perturbation can be regarded as an external force acting on the fluid in flat space~\cite{0806.0006}. Consider the covariant Navier-Stokes equations
\begin{equation}\label{ns eq}
\nabla^{\mu}T_{\mu\nu}=0,
\end{equation}
where $\nabla^{\mu}$ is compatible with $g_{\mu\nu}$. For a weakly curved spacetime, $g_{\mu\nu}=\eta_{\mu\nu}+h_{\mu\nu}$, where $\eta_{\mu\nu}$ is a Minkowski metric. For the rest of this paper we keep terms up to the first order in $h_{\mu\nu}$ only. The Navier-Stokes equation~(\ref{ns eq}) can be rewritten as
\begin{equation}
\partial^{\mu}(T_0)_{\mu\nu}=f_{\nu},
\end{equation}
where $(T_{0})_{\mu\nu}$ is stress-energy tensor of the fluid in Minkowski space. The effective forcing term $f_{\nu}$ is a  functional of $h_{\mu\nu}$ and $u_{\mu}$. Its explicit form can be found in~\cite{0806.0006}. By appropriately choosing $h_{\mu\nu}$ one can stir the fluid into various flows.

We now present the main results of this paper, before embarking on their derivation. The fluid energy density is related to pressure via equation of state
\begin{equation}\label{e+p}
\varepsilon=3P=3 (\pi T)^4,
\end{equation}
consistent with the tracelessness condition $T_{\mu}^{\mu}=0$. The tensor $\Pi_{\mu\nu}$ is
\begin{equation}\label{diss}
\begin{split}
\Pi_{\mu\nu}=&-2\eta\nabla_{\mu}u_{\nu}-\zeta \nabla_{\mu}\nabla_{\nu}\nabla u+\kappa u^{\alpha}u^{\beta}C_{\mu\alpha\nu\beta}+ \rho u^{\alpha}\nabla^{\beta} C_{\mu\alpha\nu\beta}\\
&+\xi \nabla^{\alpha}\nabla^{\beta}C_{\mu\alpha\nu\beta}- \theta u^{\alpha}\nabla_{\alpha}R_{\mu\nu},
\end{split}
\end{equation}
where $C_{\mu\alpha\nu\beta}$ and $R_{\mu\nu}$ are the Weyl and Ricci tensors of $g_{\mu\nu}$.
Appendix~\ref{appendix1} (eq.~(\ref{tensor strucutres})) provides expressions for these tensors in terms of the metric derivatives, assembled into useful combinations listed in Table~\ref{table1}. The transport coefficients are functions of derivative operators
\begin{equation}
\begin{split}
&\eta\left[u^{\alpha}\nabla_{\alpha}, \nabla_{\alpha}\nabla^{\alpha} \right], ~ \zeta\left[u^{\alpha} \nabla_{\alpha},\nabla_{\alpha} \nabla^{\alpha}\right],~
\kappa\left[u^{\alpha} \nabla_{\alpha},\nabla_{\alpha} \nabla^{\alpha}\right],\\
&\rho\left[u^{\alpha} \nabla_{\alpha},\nabla_{\alpha} \nabla^{\alpha}\right],~
\xi\left[u^{\alpha} \nabla_{\alpha},\nabla_{\alpha} \nabla^{\alpha}\right],~
\theta\left[u^{\alpha} \nabla_{\alpha},\nabla_{\alpha} \nabla^{\alpha}\right].
\nonumber
\end{split}
\end{equation}
In  Fourier space, via the replacement $\partial_{\mu}\longrightarrow\left(-i\omega,q_i\right)$, these operators become functions of $\omega$ and $q^2$. Below, we will be frequently using mixed notations: the tensor structures $\nabla_{\mu}u_{\nu}$ etc. are usually written explicitly in the form of derivatives while the transport coefficients will be be always presented as functions of momenta. The constitutive relation~(\ref{diss}) was introduced in~\cite{0905.4069}, but with $\zeta$- and $\theta$-terms missed.

The reader might be puzzled by the fact that the constitutive relation~(\ref{diss}) contains infinitely many time derivatives. As a result, the all-order resummed hydrodynamics at hand does not seem to be solvable as initial value problem, because infinitely many initial conditions would need to be supplied. To some extent this is indeed true. Nevertheless we can make sense of it by transforming the gradient expansion of the transport coefficients into memory functions with a well-prescribed initial value setup. In section~\ref{section2}, we illustrate our idea on a simple model of diffusion in magnetization.

The viscosity functions $\eta$ and $\zeta$ were computed in~\cite{1406.7222,1409.3095}, and the main results will be quoted below. The prime focus of the present paper will be on the rest of the transport coefficients that couple the stress tensor to the curvatures. In~\cite{0905.4069} these coefficients were named gravitational susceptibility of the fluid (GSFs). GSFs measure fluid's response to external gravitational perturbations. In the hydrodynamic limit, $\omega\rightarrow0$ and $q\rightarrow 0$, the viscosities and the GSFs can be expanded in powers of momenta,
and we were able to compute a few terms in the expansions analytically
\begin{equation}\label{emt expansion}
\begin{split}
&\eta=1+\frac{1}{2}\left(2-\ln{2}\right)i\omega-\frac{1}{8}q^2-\frac{1}{48}\left[6\pi- \pi^2+ 12\left(2-3\ln{2}+\ln^2{2}\right)\right]\omega^2+\cdots,\\
&\zeta=\frac{1}{12}\left(5-\pi-2\ln{2}\right)+\cdots,\\
&\kappa=2+\frac{1}{4}\left(5+\pi-6\log{2}\right)i\omega+\cdots,~~~
\rho=2+\cdots,~~~\xi=\ln{2}-\frac{1}{2}+\cdots,~~~\theta=\frac{3}{2}\zeta.
\end{split}
\end{equation}
The results are presented in terms of dimensionless frequency $\omega$ and spatial momenta $q_i$. For the rest of the paper the choice of units is set by $\pi T=1$. The physical momenta should be understood as $\pi T \omega$ and $\pi T q_i$. The first term in $\kappa$ was computed in~\cite{0712.2451,0809.4272}. $\rho$ was introduced in~\cite{0905.4069} and agrees with the result here. The relation $\theta=3\zeta/2$ holds for generic values of momenta. We complete the study of the GSFs by evaluating them numerically up to very large values of momenta and report relevant results in subsection~\ref{subsection43}. We also present a time dependent memory function derived from the shear viscosity function $\eta$. For convenience of possible applications, all our numerical results for the transport coefficient functions are deposited into a Mathematica file downloadable from the link~\cite{data}.

The technical parts of this paper are presented in sections~\ref{section3} and~\ref{section4}, which follow the setup of~\cite{1406.7222,1409.3095}. To facilitate reading of these parts, we now briefly sketch the main steps of our calculation. In section~\ref{section3}, we deal with the gravitational dynamics of a weakly perturbed  boosted black brane in AdS$_5$. The perturbation is set to be driven by varying boundary velocity field $u_\mu(x)$ and temperature $T(x)$. The boundary conditions on the 5D bulk metric are imposed so that the boundary metric is $\eta_{\mu\nu}+h_{\mu\nu}$. We then parameterize the bulk metric perturbation in terms of ten functions $h$, $k$, $j_i$ and $\alpha_{ij}$, which are both functions of the holographic coordinate and functionals of $u_\mu$ and $h_{\mu\nu}$. We then holographically read off the boundary stress tensor and express it in terms of, at this stage yet unknown, near-boundary asymptotic behavior of $h$, $k$, $j_i$ and $\alpha_{ij}$.

In section~\ref{section4}, we solve the Einstein equations for $h$, $k$, $j_i$ and $\alpha_{ij}$. This is done in two steps. First, from the vector $u_\mu$ and tensor
$h_{\mu\nu}$ we construct all possible linear structures (scalars, vectors, and tensors), which are summarized in Table~\ref{table1}. These structures serve as a basis for linear decomposition of $h$, $k$, $j_i$ and $\alpha_{ij}$. The coefficients are generally denoted as $V_i$ and $T_i$, which depend on four-momenta and the holographic coordinate only. What we achieve is that the functional dependence on unknown  velocity field and boundary metric is completely removed, and we are left with ordinary second order differential equations for $V_i$ and $T_i$. Second, we solve these equations. The results for the viscosities and the GSFs are obtained from a pre-asymptotic, near the boundary  behavior of $V_i$ and $T_i$. Summary and discussion can be found in section~\ref{section5}. We deposited many computational details into several Appendices.

\section{All order resummed hydrodynamics}\label{section2}

In this section we provide a clarification about what we actually mean by all-order resummed hydrodynamics. As has been mentioned in the Introduction, in the constitutive relation for the dissipation tensor we resumm all orders in the gradient expansion, including infinite number of time derivatives. This particularly means the effective dynamical equations in principle require an infinite set of initial conditions or, equivalently, the hydrodynamics at hand is a theory of infinite number of degrees of freedom. Indeed, these are the quasi-normal modes of the dual gravitational theory.  An alternative way to see the origin of the problem is to realize that exact gravitational theory in the bulk cannot be mapped onto a single degree of freedom on the boundary.  Below we are to solve for the bulk perturbations as a boundary value problem, with boundary conditions imposed both at the black hole horizon and the AdS boundary. If, to the contrary, we were to solve the bulk dynamics as initial value gravitational problem, we would have to supply initial conditions for the gravitational perturbation, but in  entire bulk. Roughly speaking, the initial condition along the extra dimension (holographic coordinate) would have to be mapped into infinitely many initial conditions on the boundary.

Yet, it turns out that it is possible to formulate this all-order hydrodynamics as a normal initial value problem with all the time derivatives absorbed into a complicated memory function. We illustrate our formalism by focusing on a classic problem of spin magnetization in external field \cite{km,fluid2}, which is free of unnecessary complexities related to relativistic formalism and tensorial structures, but otherwise conceptually identical to the hydrodynamics we would like to study.

Consider spin magnetization (in some direction) $M(t,\vec{x})$, which can be created by an external magnetic field $H(t,\vec{x})$. We also define the magnetization current $\vec J(t,\vec{x})$. Compared with the hydrodynamics, $M$ is identified with the velocity $u_{\mu}$, $H$ plays the same role as the external gravitational field, and $J$ is analogous to $\Pi_{ij}$. Similarly to our hydrodynamic construction, for the current we introduce the constitutive relation
\begin{equation}\label{J}
\vec J(t,\vec{x})= D(\partial_t,\nabla^2) \vec\nabla M(t,\vec{x})\,+\,\vec \sigma(\partial_t,\nabla^2)\, H(t,\vec{x}).
\end{equation}
Just like the viscosities and GSFs, the diffusion coefficient $D$ and the magnetic
susceptibility $\sigma$ are considered to be functionals of time and space derivatives. In Fourier space this reads
\begin{equation}\label{constrel}
J_i(\omega,q)= D(\omega,q^2) \,q_i\,M(\omega,q^2)\,+\,\sigma_i(\omega,q^2) \,H(\omega,q^2).
 \end{equation}
The continuity equation
\begin{equation}\label{continuity}
\dot{M}+\vec \nabla\cdot \vec J=0
\end{equation}
defines the dynamics of the system just like the conservation law for the energy momentum tensor while the constitutive relation~(\ref{constrel}) is analogous to~(\ref{diss}). Fourier transforming~(\ref{constrel}) back, we can have a different representation of~(\ref{J})
(suppressing the spatial coordinates):
\begin{equation}\label{J1}
\vec J(t)= \int_{-\infty}^\infty dt^\prime \left[ \tilde D(t-t^\prime) \vec\nabla M(t^\prime)\,+\,\tilde \sigma(t-t^\prime)\, H(t^\prime)\right].
\end{equation}
Here
\begin{equation}\label{memory}
\tilde D(t-t^\prime)\equiv\int {d\omega\over 2\pi} \,D(\omega)\,e^{-i\omega(t-t^\prime)}\,; \ \ \ \ \ \ \ \ \ \ \
\tilde \sigma(t-t^\prime)\,\equiv\int {d\omega\over 2\pi} \,\vec \sigma (\omega)\,e^{-i\omega(t-t^\prime)}.
 \end{equation}
The current $J$ at time $t$ should be affected only by the state of the system in the past. This causality requirement implies that both response functions have no support in the future: $\tilde D(t)\sim\Theta(t)$, and $\tilde \sigma(t)\sim\Theta(t)$. This is achieved only when the poles of $D(\omega)$ and $\sigma(\omega)$ are all lying below the real axis. This must be the case for any causal theory. Thus
\begin{equation}
\vec J(t)= \int_{-\infty}^t dt^\prime \left[ \tilde D(t-t^\prime) \vec\nabla M(t^\prime)\,+\,\tilde \sigma(t-t^\prime)\, H(t^\prime)\right].
\end{equation}

The external magnetic field $H$ is normally turned on at negative times, so  to create initial magnetization at $t=0$, and then turned off at $t=0$ ($H(t>0)=0$), letting the system to freely relax to its equilibrium at infinite future. For such experimental setup, for positive times the current
\begin{equation}\label{J2}
\vec J(t>0)= \int_{0}^t dt^\prime \tilde D(t-t^\prime) \vec\nabla M(t^\prime)\,+\,\vec J_H(t)
 \end{equation}
with
\begin{equation}\label{JH}
\vec J_H(t)= \int_{-\infty}^0 dt^\prime \left[ \tilde D(t-t^\prime) \vec\nabla M(t^\prime)\,+\,\tilde \sigma(t-t^\prime)\,H(t^\prime)\right].
 \end{equation}
Generically, $J_H$ is not vanishing and it accounts for the entire history of the system at negative times. This is in contrast to a typical memory function-based approach, where one introduces constitutive relation~(\ref{J2}) assuming $J_H=0$ and then also models
$\tilde D$ \cite{km,fluid2}.

Our construction is so far formally exact. However, in order to solve the dynamical
equation~(\ref{continuity}), it is not sufficient to provide the initial condition for magnetization only, but we also need the ``history'' current $J_H$ at all times, equivalent to providing infinitely many additional initial conditions.

We are now to discuss under what conditions we can nevertheless set $J_H$ to zero, casting our theory into a well-defined initial value problem. The response functions $\tilde D$ and $\tilde \sigma$ are some given functions defined by underlying microscopic theory. Thus the equation $J_H(t>0)=0$ is in fact an equation for the magnetic field $H$
at $t<0$. It is, however, not obvious that there exists a solution for generic $\tilde D$ and $\tilde \sigma$, because we want the current $J_H=0$ at all positive times. Even though we cannot guarantee vanishing of the current identically, it is safe to assume that its effect could be rendered negligibly small. Particularly, it is obvious that $J_H$ vanishes at late times and its only potential influence could be at very early times when the current $J\simeq J_H$.

When it comes to modeling of $\tilde D$, a first try is usually the relaxation time approximation~\cite{Muller,Israel,IS1976,IS1979}:
\begin{equation}
\tilde D_{IS}(t)\,=\,D_0\, e^{-t/\tau_R}
\end{equation}
This model assumes $D(\omega)$ has a purely imaginary and momentum independent pole at $\omega=-i/\tau_R$. A further improvement of this model would be to account for additional poles, say, $\omega_{\pm}=\pm\omega_R-i\omega_I$~\cite{1409.5087}: \footnote{With an equivalent model for the viscosity function, we almost recover eqs.~(12),(15),(16) of~\cite{1409.5087}. However, our version of eq.~(12) has an extra term and a source term.}
\begin{equation}
\tilde D_{\pm}(t)\,=\,d_0\, e^{-t/\tau_R}\,+\,d_1\, e^{-t\,\omega_I}\,\cos(\omega_R\,t)
\end{equation}
In contrast to the above models, a calculation analogous to one presented below for the viscosity functions and the GSFs, would determine $D$ and $\sigma$ exactly, from the underlying microscopic theory. Any practical application of thus resumed hydrodynamics  would have to additionally impose $J_H=0$.

After this explanatory section, we return to the main bulk of our work, and that is to  determine the viscosity functions and the GSFs for ${\cal N}=4$ CFT precisely.

\section{Linearized fluid/gravity correspondence: weakly curved boundary} \label{section3}
\paragraph{Notations and Conventions.} Upper case Latin indices $\left\{M,N,\cdots\right\}$ and lower case Greek indices $\left\{\mu,\nu,\cdots\right\}$
denote the bulk and boundary directions, respectively. Lower case Latin indices $\left\{i,j,\cdots\right\}$ are used to specify spatial directions on the boundary. Indices $v$ and $0$ will be used alternatively to denote the time direction. We take the $\textrm{AdS}_5$ radius to be unity which is equivalent to setting the bulk cosmological constant $\Lambda=-6$. The bulk metric is $G_{MN}$, and $g_{\mu\nu}$ denotes (non-dynamical) metric in the boundary theory. Throughout this paper, we work with the mostly plus signature for the metric tensors. Further details can be found in Appendix~\ref{appendix1}.

A universal subsector of the AdS/CFT correspondence is the Einstein-Hilbert term plus a negative cosmological constant,
\begin{equation}\label{eh action}
S_{\textrm{EH}}=\frac{1}{16\pi G_N}\int d^5x\sqrt{-G}\left(R+12\right).
\end{equation}
Its variation leads to Einstein equations in the bulk,
\begin{equation}\label{einstein eq}
E_{MN}=R_{MN}-\frac{1}{2}G_{MN}R-6G_{MN}=0.
\end{equation}
For the variational principle to be well-defined, the action~(\ref{eh action}) has to be supplemented with the Gibbons-Hawking surface term
\begin{equation}\label{gh action}
S_{\text{GH}}=\frac{1}{8\pi G_N}\int d^4x\sqrt{-\gamma}~K[\gamma]
\end{equation}
where $\gamma_{\mu\nu}$ is an induced metric on a fixed $r$ hypersurface $\Sigma$ with $r$ being the holographically emergent coordinate. The extrinsic curvature $K_{\mu\nu}$ of $\Sigma$ is
\begin{equation}
K_{\mu\nu}=\gamma_{\mu}^{\alpha}\bar{\nabla}_{\alpha}n_{\nu},
\end{equation}
where $n_{\mu}$ is an outgoing vector, normal to $\Sigma$ and $\bar{\nabla}$ is compatible with $\gamma_{\mu\nu}$. The boundary metric $g_{\mu\nu}$ is related to $\gamma_{\mu\nu}$ through $g_{\mu\nu}=\lim_{r\to\infty}\left(\gamma_{\mu\nu}/r^2\right)$.

To ensure finiteness of the boundary stress-energy tensor, we use  holographic renormalisation and add to the actions~(\ref{eh action},\ref{gh action}) a suitable counter-term $S_{\textrm{c.t.}}$. For curved boundary relevant to our study, the counter-term was worked out in~\cite{hep-th/0002230,hep-th/0010138}
\begin{equation}\label{ct action}
S_{\textrm{c.t.}}=-\frac{1}{16\pi G_N}\int d^4x\sqrt{-\gamma} \left\{6+\frac{1}{2} R[\gamma]-\log\frac{1}{r^2} \left(\frac{1}{8} R^{\mu\nu}[\gamma] R_{\mu\nu}[\gamma] -\frac{1}{24}R^2 [\gamma]\right)\right\},
\end{equation}
which is a functional of the induced metric $\gamma_{\mu\nu}$. The boundary stress-energy tensor $T_{\mu\nu}$ is defined as follows
\begin{equation}\label{st def}
T_{\mu\nu}=\lim_{r\to \infty}\tilde{T}_{\mu\nu}(r)\equiv-\lim_{r\to\infty} \left(r^2\cdot \frac{2}{\sqrt{-\gamma}}\frac{\delta S_{\textrm{ren}}} {\delta\gamma^{\mu\nu}}\right),
\end{equation}
where the renormalized action $S_\textrm{ren}$ is
\begin{equation}
S_{\textrm{ren}}=S_{\textrm{EH}}+S_{\textrm{GH}}+S_{\text{c.t.}}.
\end{equation}
From the definition~(\ref{st def}), we have
\begin{equation}\label{stress tensor}
\tilde{T}_{\mu\nu}(r)=\underbrace{-2r^2\left(K_{\mu\nu}-K\gamma_{\mu\nu}+ 3\gamma_{\mu\nu}- \frac{1}{2} \mathcal{G}_{\mu\nu}(\gamma) \right)}_{\tilde{T}_{\mu\nu}^{n}(r)} +\underbrace{2\,T_{\mu\nu}^{a}\log\frac{1}{r^2}}_{\tilde{T}_{\mu\nu}^{a}(r)},
\end{equation}
where $\mathcal{G}_{\mu\nu}(\gamma)$ is the Einstein tensor compatible with $\gamma_{\mu\nu}$. The expression for $T_{\mu\nu}^a$ can be found in~\cite{hep-th/0002230}\footnote{Due to differences in signature convention of the metric tensor, the Riemann curvature of~\cite{hep-th/0002230} has an additional overall minus sign.}
\begin{equation}\label{stress tensor a}
\begin{split}
T_{\mu\nu}^a=&-\frac{1}{4}R_{\mu\rho\nu\lambda}[g]R^{\rho\lambda}[g]+\frac{1}{24} \nabla_{\mu} \nabla_{\nu}R[g]-\frac{1}{8}\nabla^2R_{\mu\nu}[g]+\frac{1}{12}R[g] R_{\mu\nu}[g]\\
&+\frac{1}{48}g_{\mu\nu}\left( \nabla^2R[g]-R^2[g]+3R_{\rho\lambda}[g] R^{\rho\lambda} [g]\right).
\end{split}
\end{equation}
When the boundary is flat, $g_{\mu\nu}\rightarrow\eta_{\mu\nu}$, $\tilde T^a_{\mu\nu}$ vanishes and we are left with $\tilde{T}_{\mu\nu}^{n}$, which was first worked out in~\cite{hep-th/9902121}. The second piece, $\tilde{T}_{\mu\nu}^a$, is crucial in removing logarithmic divergences, which emerge in $\tilde T^n_{\mu\nu}$ due to boundary curvature. Finally, in~(\ref{stress tensor}) we normalized the Newton constant $G_{N}$ by $16\pi G_{N}=1$ so that $\varepsilon$ and $P$ take the forms~(\ref{e+p}).

In order to incorporate metric perturbation in the boundary theory, we follow~\cite{0806.0006,0809.4272} and consider the boosted asymptotically locally $\textrm{AdS}_5$ black brane metric,
\begin{equation}\label{metric1}
d\mathfrak{s}^2=-2u_{\mu}(x^\alpha)dx^{\mu}dr-r^2f\left({\bf b}(x^\alpha) r\right) u_{\mu}(x^\alpha)u_{\nu}(x^\alpha)dx^{\mu}dx^{\nu}+r^2 \Delta_{\mu\nu}(x^\alpha) dx^{\mu}dx^{\nu},
\end{equation}
where $f(r)=1-1/r^4$. Hence $r=1$ is the horizon. The conformal boundary is at $r=\infty$. The velocity field $u_{\mu}(x^{\alpha})$ is normalized in a standard way
\begin{equation}\label{normalization}
g^{\mu\nu}(x^{\alpha})u_{\mu}(x^{\alpha})u_{\nu}(x^{\alpha})=-1.
\end{equation}
The temperature field $T(x^{\alpha})$ is related to the parameter ${\bf b}(x^{\alpha})$ via
\begin{equation}
T(x^{\alpha})=\frac{1}{\pi {\bf b}(x^{\alpha})},
\end{equation}
which is identified as temperature of the boundary CFT. If $u_{\mu}(x^{\alpha})$, ${\bf b}(x^{\alpha})$ and $g_{\mu\nu}(x^{\alpha})$ are taken constant
\begin{equation}
u_{\mu}(x^{\alpha})\dashrightarrow u_{\mu},~~~{\bf b}(x^{\alpha})\dashrightarrow {\bf b},~~~g_{\mu\nu}(x^{\alpha})\dashrightarrow \eta_{\mu\nu},
\end{equation}
the line element~(\ref{metric1}) does form a class of solutions to the bulk Einstein equations~(\ref{einstein eq}). As $r\to\infty$, the metric~(\ref{metric1}) approaches asymptotics
\begin{equation}
d\mathfrak{s}^2\xlongrightarrow{r\to\infty}-2u_{\mu}(x^\alpha)dx^{\mu}dr+r^2 g_{\mu\nu}(x^\alpha) dx^{\mu}dx^{\nu}+\mathcal{O}\left(\frac{1}{r^2}\right),
\end{equation}
which identifies $g_{\mu\nu}(x^{\alpha})$ as the background metric for the boundary theory.

To study fluid dynamics we are to consider velocity and temperature as given, but at this stage arbitrary functions of boundary coordinates. Then the element~(\ref{metric1}) ceases to solve the Einstein equations. In order to find a solution, we have to add  to~(\ref{metric1}) a correction which is to be determined by solving the Einstein equations~(\ref{einstein eq}). Our main goal is to compute the stress-energy tensor with all the derivative terms linear in the fluid dynamical variables and boundary metric perturbation resummed. These variables can be expanded to linear order
\begin{equation}\label{linearization}
\begin{split}
u_{\mu}(x^{\alpha})&=\left(-1+\frac{1}{2}\epsilon h_{00}(x^{\alpha}),\epsilon u_{i} (x^{\alpha}) \right)+\mathcal{O}\left(\epsilon^2\right),\\
g_{\mu\nu}(x^{\alpha})&=\eta_{\mu\nu}+\epsilon h_{\mu\nu}(x^{\alpha})+\mathcal{O} (\epsilon^2),\\
{\bf b}(x^{\alpha})&={\bf b}_0+\epsilon {\bf b}_1(x^{\alpha})+\mathcal{O} (\epsilon^2),
\end{split}
\end{equation}
where $\epsilon$ is an order counting parameter and will be set to one in the end. As seen from~(\ref{linearization}), gravitational perturbations can induce fluid flow\footnote{Recently, an opposite effect, generation of gravitational waves from sounds was considered in~\cite{1304.2433,1412.5147}.}. The constant ${\bf b}_0$ corresponds to the equilibrium temperature of the boundary theory and will be set to one from now on. The seed metric, i.e., a linearized version of~(\ref{metric1}) is
\begin{equation}\label{seed}
\begin{split}
ds^2_{\text{seed}}=&\,2drdv-r^2f(r)dv^2+r^2d{\vec{x}}^2\\
&-\epsilon\left[2u_i(x^{\alpha}) drdx^i+\frac{2}{r^2}u_i(x^{\alpha}) dvdx^i+ \frac{4}{r^2} {\bf b}_1(x^{\alpha})dv^2\right.\\
&\left.+h_{00}(x^\alpha)drdv+\frac{1}{r^2}h_{00}(x^\alpha)dv^2-r^2h_{\mu\nu}(x^{\alpha}) dx^\mu dx^\nu\right]+\mathcal{O}(\epsilon^2).
\end{split}
\end{equation}
Denoting the metric correction as $ds^2_{\textrm{corr}}$, the full metric is then formally written as
\begin{equation}
ds^2=G_{MN}dx^{M}dx^{N}=ds^2_{\text{seed}}+ds^2_{\text{corr}}.
\end{equation}
For the metric correction we choose the ``background field'' gauge~\cite{0712.2456}
\begin{equation}\label{bk gauge}
G_{rr}=0,~~~G_{r\mu}\propto u_{\mu},~~~\textrm{Tr}\left[(G^{(0)})^{-1}G^{(1)}\right]=0,
\end{equation}
where $G^{(0)}$ corresponds to the first line of~(\ref{seed}) and $G^{(1)}$ stands for the metric correction $ds_{\textrm{corr}}^2$.
Then the most general form of $ds^2_{\text{corr}}$ can be cast into the following form
\begin{equation}\label{line element correction}
ds^2_{\text{corr}}=\epsilon\left(-3hdrdv+\frac{k}{r^2}dv^2 +r^2 h d{\vec{x}}^2+ \frac{2}{r^2} j_{i}dvdx^i+r^2\alpha_{ij}dx^idx^j\right),
\end{equation}
where $\alpha_{ij}$ is a traceless symmetric tensor of rank two. We would like to stress that all the components $\{h,~k,~j_{i},~\alpha_{ij}\}$ are explicit functions of the bulk coordinates $\{x^{\alpha},r\}$ and, through the Einstein equations, also become functionals of $u_i(x^{\alpha})$ and $h_{\mu\nu}(x^{\alpha})$.

In order to solve for the metric correction, we have to impose appropriate boundary conditions. The relevant ones have been discussed in details in~\cite{1406.7222,1409.3095}. For completeness, we briefly summarize them here. The first type is a regularity requirement for all the components $\{h,~k,~j_{i},~\alpha_{ij}\}$ over the whole range of $r$. Second, since the boundary theory is supposed to live in a $\emph{fixed}$ spacetime with given metric $g_{\mu\nu}(x^{\alpha})$, near the boundary $r=\infty$ we require
\begin{equation}\label{AdS constraint}
h< \mathcal{O}(r^0),~~~k<\mathcal{O}(r^4),~~~j_i<\mathcal{O}(r^4),~~~ \alpha_{ij}<\mathcal{O}(r^0).
\end{equation}
Finally, the remaining ambiguity is fixed by defining the fluid velocity $u_{\mu}(x^{\alpha})$ in Landau frame,
\begin{equation}
u^{\mu}T_{\mu\nu}=-\varepsilon u_{\nu}\Longrightarrow u^{\mu}\Pi_{\langle\mu\nu\rangle} =0
\end{equation}
which, within the linear approximation, is equivalent to specifying $\Pi_{\langle\mu\nu\rangle}$ as transverse
\begin{equation}\label{frame convention}
\Pi_{\langle00\rangle}=\Pi_{\langle0i\rangle}=0,~~\Pi_{\langle ij\rangle}\neq 0.
\end{equation}
In Appendix~\ref{appendix2}, the tensor $\tilde{T}_{\mu\nu}(r)$ is explicitly expressed in terms of the metric components~(\ref{line element correction}). So, once the metric corrections are found from the Einstein equations, it is straightforward to derive the stress-energy tensor from~(\ref{stress tensor1},\ref{stress tensor2}).

\section{From gravity to  fluid dynamics}\label{section4}
In this section, we derive the boundary fluid dynamics through solving the gravitational equations in the bulk. The expressions~(\ref{einstein eq}) are a set of fifteen equations, with one redundant component. The remaining equations can be split into ten dynamical equations and four constraints. The dynamical equations are to be solved  for $h$, $k$, $j_i$ and $\alpha_{ij}$, which are precisely ten unknown  functions in the metric corrections~(\ref{line element correction}). Thus obtained solutions would still functionally depend on the, yet unspecified, velocity field and background metric.  For  the boundary fluid, solving for dynamical components of~(\ref{einstein eq}) would mean that  we first construct an ``off-shell'' stress-energy tensor.  It is, however, worth emphasizing that this procedure is sufficient to uniquely fix the transport coefficient functions.

The remaining four constraint equations act as constraints in the space of velocity and temperature fields, for which solutions to the gravitational equations exist. These equations are found to coincide with the Navier-Stokes equations for the resummed hydrodynamics. In other words, satisfying the constraints is equivalent to putting the stress-energy tensor ``on-shell''. We demonstrate this equivalence in Appendix~\ref{appendix4}, where the conservation laws of $T_{\mu\nu}$ are shown be to consistent with the constraints.

\subsection{Dynamical Einstein equations}\label{subsection41}

The first dynamical equation is $E_{rr}=0$, which yields
\begin{equation}\label{h eq}
5\partial_rh+r\partial_r^2h=0.
\end{equation}
The asymptotic requirement~(\ref{AdS constraint}) and the fluid frame convention $\Pi_{\langle 00\rangle}=0$ lead to $h=0$. The function $k$ can be found
from $E_{rv}=0$,
\begin{equation}\label{k eq}
\begin{split}
\partial_rk=&2r^2\partial u-2r^2\partial_kh_{0k}-\frac{1}{6}r\partial^2h_{00}-  \frac{1}{3} r \left(\partial_i\partial_jh_{ij}-\partial^2h_{kk}\right)\\
&+r^2\partial_vh_{kk}+\frac{1}{3}r\partial_{v}\partial u-\frac{2}{3r^2}\partial j
-\frac{1}{3}r\partial_i\partial_j\alpha_{ij}-\frac{1}{3r}\partial_r\partial j,
\end{split}
\end{equation}
which is coupled to $j_i$ and $\alpha_{ij}$. Our strategy is the same as that in~\cite{1409.3095}: we will first solve for  the functions $j_i$ and $\alpha_{ij}$,
and then integrate  the equation  for $k$.

The equation for $j_i$ is derived  from $E_{ri}=0$,
\begin{equation}\label{j eq}
0=r\partial_r^2j_i-3\partial_rj_i+3r^2\partial_vu_i+r\left(\partial^2u_i-\partial_i \partial u\right)-\frac{3}{2}r^2\partial_ih_{00}+r^3\partial_r\partial_j\alpha_{ij},
\end{equation}
which is  coupled with $\alpha_{ij}$. The equation for the latter is determined from $E_{ij}=0$, after some massaging, which is presented for the flat space in~\cite{1409.3095}. Extension to a curved space is straightforward and here we quote the final equation only
\begin{equation}\label{alpha eq}
\begin{split}
0=&(r^7-r^3)\partial_r^2\alpha_{ij}+(5r^6-r^2)\partial_r\alpha_{ij}+2r^5\partial_r \partial_v \alpha_{ij}+3r^4\partial_v\alpha_{ij}\\
&+r^3\left(\partial^2\alpha_{ij}-\partial_i\partial_k\alpha_{ik}-\partial_j\partial_k \alpha_{ik} +\frac{2}{3}\delta_{ij}\partial_k\partial_l\alpha_{kl}\right)\\
&+(1-r\partial_r)\left(\partial_ij_j+\partial_jj_i-\frac{2}{3}\delta_{ij}\partial j\right)+(r^3\partial_v+3r^4)\left(\partial_iu_j+\partial_ju_i-\frac{2}{3}\delta_{ij} \partial u\right)\\
&+3r^4\partial_v\left(h_{ij}-\frac{1}{3}\delta_{ij}h_{kk}\right)-r^3\left(\partial_i \partial_j h_{00}-\frac{1}{3}\delta_{ij}\partial^2\partial^2 h_{00}\right)\\
&-3r^4\left(\partial_ih_{0j}+\partial_{j}h_{0i}-\frac{2}{3}\delta_{ij}\partial_{k}h_{0k} \right)\\
&+r^3\left(\partial^2h_{ij}-\partial_i\partial_kh_{jk}-\partial_j\partial_kh_{ik}+ \partial_i \partial_j h_{kk}-\frac{2}{3}\delta_{ij}\partial^2h_{kk}+\frac{2}{3}\delta_{ij}\partial_k\partial_l h_{kl}\right).
\end{split}
\end{equation}

Eqs.~(\ref{k eq}, \ref{j eq}, \ref{alpha eq}) are partial differential equations. Furthermore, they all have source terms, which depend on the velocity field $u_i$ and background metric $h_{\mu\nu}$. Those are arbitrary at this moment. Thus, solutions of~(\ref{k eq}, \ref{j eq}, \ref{alpha eq}) are functionals of $u_i$ and  $h_{\mu\nu}$. Our strategy for solving these equations goes in two steps. First, due to the linear approximation, we Fourier transform these equations  with respect to the boundary coordinates. In this way we cast these equations into ordinary differential equations
with respect to the holographic coordinate $r$. As a second step, we rid off the functional dependence on $u_i$ and  $h_{\mu\nu}$.  This is done by decomposing the vector $j_i$ and tensor $\alpha_{ij}$  using base tensor structures constructed from $u_i$ and $h_{\mu\nu}$. These structures are classified according to $SO(3)$ symmetry and are listed in Table~\ref{table1}. We end up with a large system of (partially coupled) ordinary differential equations for the coefficient functions.
\begin{table}[htbp]
\tabcolsep 6mm
\begin{center}
\begin{tabular}{| l | l | l | }
\hline
Scalar & Vector &~~~~~~~~~~~~~~~~~~~~~Tensor \\
\hline
${\bf{b}}_1$ &$u_i$&$\tau_{ij}\equiv h_{ij}-\frac{1}{3}\delta_{ij}h_{kk}$\\
$h_{00}$ & $h_{0i}$ &$\hat{\sigma}_{ij}\equiv\frac{1}{2}\left(\partial_iu_j+\partial_ju_i- \frac{2}{3}\delta_{ij} \partial u\right)$  \\
$h_{kk}$ &$\partial_i{\bf{b}}_1$  & $\tilde{\sigma}_{ij}\equiv\frac{1}{2}\left(\partial_ih_{0j}+\partial_jh_{0i}-\frac{2}{3} \delta_{ij}\partial_k h_{0k}\right)$ \\
$\partial u$& $\partial_i h_{00}$ &$\varpi_{ij}\equiv\partial_i\partial_j {\bf{b}}_1- \frac{1}{3}\delta_{ij}\partial^2{\bf{b}}_1$\\
$\partial_k h_{0k}$&$\partial_ih_{kk}$ & $\varphi_{ij}\equiv\partial_i\partial_j h_{00}- \frac{1}{3} \delta_{ij}\partial^2 h_{00}$\\
$\partial_i\partial_j h_{ij}$&$\partial_i\partial u$&$\chi_{ij}\equiv\partial_i\partial_j h_{kk}- \frac{1}{3} \delta_{ij}\partial^2 h_{kk}$\\
&$\partial_i\partial_k h_{0k}$&$\psi_{ij}\equiv\frac{1}{2} \left(\partial_i\partial_k h_{jk}+ \partial_j \partial_k h_{ik}- \frac{2}{3} \delta_{ij} \partial_k\partial_l h_{kl} \right)$\\
&$\partial_i\partial_k\partial_l h_{kl}$&$\hat{\pi}_{ij}\equiv\partial_i\partial_j \partial u-\frac{1}{3}\delta_{ij}\partial^2\partial u$\\
&&$\tilde{\pi}_{ij}\equiv\partial_i\partial_j\partial_k h_{0k}-\frac{1}{3}\delta_{ij} \partial^2 \partial_kh_{0k}$\\
&&$\phi_{ij}\equiv\partial_i\partial_j\partial_k\partial_lh_{kl}-\frac{1}{3}\delta_{ij} \partial^2\partial_k\partial_l h_{kl}$\\
\hline
\end{tabular}
\end{center}
\caption{Up to linear order $\mathcal{O}(\epsilon)$, we exhaustively list all the basic tensor structures constructed from the fluid dynamic variables $u_{\mu}$, ${\bf{b}}_1$ and $g_{\mu\nu}$, classified according to the $SO(3)$ symmetry.}\label{table1}
\end{table}

To proceed, $j_i$ and $\alpha_{ij}$ are uniquely decomposed as
\begin{align}\label{decomposition}
j_i&=V_1 u_i+V_2\partial_i\partial u+ V_3 h_{0i} + V_4\partial_i\partial_kh_{0k}
+ V_5\partial_ i h_{00}+ V_6 \partial_i h_{kk}+V_7 \partial_kh_{ik}+V_8 \partial_i\partial_k\partial_l h_{kl},\nonumber\\
\alpha_{ij}&=2T_1\hat{\sigma}_{ij}+T_2\hat{\pi}_{ij}+ 2T_3 \tilde{\sigma}_{ij} +T_4\tilde{\pi}_{ij} +T_5\varphi_{ij}+ T_6\chi_{ij}+ 2T_7\psi_{ij}+T_8\phi_{ij} +T_9\tau_{ij},
\end{align}
where $V_i$ and $T_i$ are short notations for $V_i(\omega,q_i,r)$ and $T_i(\omega,q_i,r)$. Since the temperature variation ${\bf b}_1$ never appears in any of the dynamical equations, the tensor structures constructed from ${\bf{b}}_1$ in Table~\ref{table1} do not enter the decomposition~(\ref{decomposition}). Substituting~(\ref{decomposition}) into eqs.~(\ref{j eq},\ref{alpha eq}), we arrive at the ordinary differential equations for $V_i$ and $T_i$. In what follows, we present the equations grouped into decoupled sectors.
\paragraph{I:~$\left\{V_1,~V_2,~T_1,~T_2\right\}$}
\begin{equation}\label{v12t12}
\left\{
\begin{aligned}
0=&r\partial_r^2V_1-3\partial_rV_1-3i\omega r^2-q^2r-q^2r^3\partial_rT_1, \\
0=&r\partial_r^2V_2-3\partial_rV_2-r+\frac{1}{3}r^3\partial_r T_1-\frac{2}{3}q^2r^3 \partial_rT_2,\\
0=&(r^7-r^3)\partial_r^2T_1+(5r^6-r^2)\partial_rT_1-2i\omega r^5\partial_rT_1\\
&-3i\omega r^4 T_1+V_1-r\partial_rV_1-i\omega r^3+3r^4,\\
0=&(r^7-r^3)\partial_r^2T_2+(5r^6-r^2)\partial_rT_2-2i\omega r^5\partial_rT_2\\
&-3i\omega r^4 T_2+2V_2-2r\partial_rV_2+\frac{1}{3}q^2r^3T_2-\frac{2}{3}r^3T_1.
\end{aligned}
\right.
\end{equation}
This first sector decouples from the metric perturbation and  was explored in~\cite{1406.7222,1409.3095}. The holographic RG-flow type equations~(\ref{v12t12}) determine the viscosity functions $\eta$ and $\zeta$.

The remaining sectors are new. They all emerge due to the metric perturbation.
\paragraph{II:~$\left\{V_3,~V_4,~T_3,~T_4\right\}$}
\begin{equation} \label{v34t34}
\left\{
\begin{aligned}
0=&r\partial_r^2V_3-3\partial_rV_3-q^2r^3\partial_rT_3, \\
0=&r\partial_r^2V_4-3\partial_rV_4+\frac{1}{3}r^3\partial_r T_3-\frac{2}{3}q^2r^3 \partial_rT_4,\\
0=&(r^7-r^3)\partial_r^2T_3+(5r^6-r^2)\partial_rT_3-2i\omega r^5\partial_rT_3\\
&-3i\omega r^4 T_3+V_3-r\partial_rV_3-3r^4,\\
0=&(r^7-r^3)\partial_r^2T_4+(5r^6-r^2)\partial_rT_4-2i\omega r^5\partial_rT_4\\
&-3i\omega r^4 T_4+2V_4-2r\partial_rV_4+\frac{1}{3}q^2r^3T_4-\frac{2}{3}r^3T_3.
\end{aligned} \right.
\end{equation}
\paragraph{III:~$\left\{V_5,~T_5\right\}$}
\begin{equation}
\left\{
\begin{aligned}
0=&r\partial_r^2V_5-3\partial_rV_5-\frac{3}{2}r^2-\frac{2}{3}q^2r^3\partial_rT_5,\\
0=&(r^7-r^3)\partial_r^2T_5+(5r^6-r^2)\partial_rT_5-2i\omega r^5\partial_rT_5\\
&-3i\omega r^4 T_5+\frac{1}{3}q^2r^3T_5+2V_5-2r\partial_rV_5-r^3.
\end{aligned}
\right.
\end{equation}
\paragraph{IV:~$\left\{T_9\right\}$}
\begin{equation}
\begin{split}
0=&(r^7-r^3)\partial_r^2T_9+(5r^6-r^2)\partial_rT_9-2i\omega r^5\partial_rT_9\\
&-3i\omega r^4 T_9-q^2r^3T_9-3i\omega r^4-q^2r^3.
\end{split}
\end{equation}
\paragraph{V:~$\left\{V_6,T_6\right\}$}
\begin{equation}
\left\{
\begin{aligned}
0=&r\partial_r^2V_6-3\partial_rV_6-\frac{1}{3}r^3\partial_rT_9-\frac{2}{3}q^2r^3 \partial_rT_6,\\
0=&(r^7-r^3)\partial_r^2T_6+(5r^6-r^2)\partial_rT_6-2i\omega r^5\partial_rT_6\\
&-3i\omega r^4T_6+2V_6-2r\partial_rV_6+\frac{1}{3}q^2r^3T_6+\frac{2}{3}r^3T_9+r^3.
\end{aligned}
\right.
\end{equation}
\paragraph{VI:~$\left\{V_7,~T_7\right\}$}
\begin{equation}
\left\{
\begin{aligned}
0=&r\partial_r^2V_7-3\partial_rV_7+r^3\partial_rT_9-q^2r^3\partial_rT_7,\\
0=&(r^7-r^3)\partial_rT_7+(5r^6-r^2)\partial_rT_7-2i\omega r^5\partial_rT_7\\
&-3i\omega r^4 T_7+V_7-r\partial_rV_7-r^3 T_9-r^3.
\end{aligned}
\right.
\end{equation}
\paragraph{VII:~$\left\{V_8,~T_8\right\}$}
\begin{equation}\label{V8}
\left\{
\begin{aligned}
0=&r\partial_r^2V_8-3\partial_rV_8+\frac{1}{3}r^3\partial_rT_7-\frac{2}{3}q^2r^3 \partial_rT_8,\\
0=&(r^7-r^3)\partial_r^2T_8+(5r^6-r^2)\partial_rT_8-2i\omega r^5\partial_r T_8\\
&-3i\omega r^4T_8+2V_8-2r\partial_rV_8+\frac{1}{3}q^2r^3T_8-\frac{2}{3}r^3T_7.
\end{aligned}
\right.
\end{equation}
The transport coefficients are completely fixed by integrating these equations from the horizon to the boundary. We, however, postpone this integration step until subsection~\ref{subsection43}. In the next subsection~\ref{subsection42}, we focus on the fluid stress-energy tensor, which so far is expressed in terms of $h$, $k$, $j_i$ and $\alpha_{ij}$ (see~(\ref{stress tensor1},\ref{stress tensor2})). Via the decomposition~(\ref{decomposition}), the stress tensor can be written in terms of momenta- and $r$-dependent coefficient functions $V_i$ and $T_i$ multiplying the tensorial structures listed in Table~\ref{table1}. Furthermore, we are mostly interested in these coefficient functions near the boundary. Eventually we will be able to map certain pre-asymptotic behavior of these coefficient functions onto momenta-dependent transport coefficients. To this goal, we first analyze the above equations near $r=\infty$.

\subsection{Stress-energy tensor from  near the boundary analysis}\label{subsection42}

We are to study the large $r$ behavior of the metric corrections, which makes it possible to cast the stress tensor into the form~(\ref{st result1},\ref{diss}), but with the transport coefficients left undetermined. The latter cannot be determined from asymptotic considerations only, but require a full integration over the holographic coordinate.

We begin with the first sector $\left\{V_1,~T_1,~V_2,~T_2\right\}$, which was already analyzed in~\cite{1409.3095}:
\begin{equation}\label{asym 12}
\begin{split}
&V_1\xlongrightarrow{r\to\infty} -i\omega r^3+\mathcal{O}\left(\frac{1}{r}\right),~~~ T_1\xlongrightarrow{r\to\infty} \frac{1}{r}+\frac{t_1}{r^4}+\mathcal{O} \left(\frac{1}{r^5}\right), \\
&V_2\xlongrightarrow{r\to\infty} -\frac{1}{3}r^2+\mathcal{O} \left(\frac{1}{r}\right), ~~~~
T_2\xlongrightarrow{r\to\infty} \frac{t_2}{r^4}+\mathcal{O}\left(\frac{1}{r^4}\right).
\end{split}
\end{equation}
The momenta-dependent coefficients $t_1$ and $t_2$ are related to the viscosities~\cite{1406.7222,1409.3095}
\begin{equation}
\eta=-4t_1,~~~~\zeta=-4t_2.
\end{equation}
In the limit  $r\to \infty$, the functions $\left\{V_3,~V_4,~T_3,~T_4\right\}$ have the following expansion
\begin{equation}\label{asym 34}
\begin{split}
&V_3\longrightarrow -\frac{1}{4}q^2r^2+\frac{1}{6}i\omega q^2 r+\frac{1}{16}q^2(q^2- \omega^2) \log{r}+\frac{1}{64}q^2(\omega^2+3q^2)+\mathcal{O} \left(\frac{\log{r}}{r} \right),\\
&T_3\longrightarrow -\frac{1}{r}+\frac{i\omega}{4r^2}+\frac{q^2-\omega^2}{12r^3}+ \frac{t_3}{r^4}-\frac{1}{16r^4}i\omega (\omega^2-q^2)\log{r}+\mathcal{O} \left(\frac{\log{r}}{r^5}\right),\\
&V_4\longrightarrow \frac{1}{12}r^2-\frac{1}{18}i\omega r-\frac{1}{192}(\omega^2-9q^2)
+\frac{\omega^2+3q^2}{48} \log{r}+\mathcal{O}\left(\frac{\log{r}}{r}\right),\\
&T_4\longrightarrow \frac{1}{6r^3}+\frac{t_4}{r^4}+\frac{i\omega \log{r}}{12r^4} +\mathcal{O}\left(\frac{\log{r}}{r^5}\right).
\end{split}
\end{equation}
As $r\to\infty$, all the other functions have the asymptotic behaviors of the form,
\begin{equation}\label{asym 56789}
\begin{split}
V_5\longrightarrow &-\frac{1}{2}r^3+\frac{1}{9}q^2r -\frac{1}{96}i\omega q^2
+\frac{1}{24}i\omega q^2 \log{r}+ \mathcal{O}\left(\frac{\log{r}}{r}\right),\\
T_5\longrightarrow &\frac{1}{4r^2}+\frac{i\omega}{12r^3}+\frac{t_5}{r^4}+\frac{1}{48r^4} (q^2-3\omega^2)\log{r} + \mathcal{O}\left(\frac{\log{r}}{r^5}\right),\\
V_6\longrightarrow &-\frac{1}{12}i\omega r^2+\frac{1}{18}(q^2-\omega^2)r -\frac{1}{192}i \omega (q^2-\omega^2)\\
&+\frac{1} {48} i\omega (q^2-\omega^2)\log{r} + \mathcal{O} \left(\frac{\log{r}}{r}\right),\\
V_7\longrightarrow &\frac{1}{4}i\omega r^2+\frac{1}{6}\omega^2r -\frac{1}{64}i\omega (\omega^2+3q^2)-\frac{1}{16}i\omega (q^2-\omega^2)\log{r}+ \mathcal{O} \left(\frac{\log{r}}{r}\right),\\
V_8\longrightarrow &\frac{1}{18}r-\frac{5}{96}i\omega
-\frac{1}{24}i\omega \log{r} + \mathcal{O} \left(\frac{\log{r}}{r}\right),\\
T_6\longrightarrow &\frac{1}{4r^2}-\frac{i\omega}{12r^3}+\frac{t_6}{r^4}-\frac{1}{48r^4} (q^2-\omega^2)\log{r} + \mathcal{O}\left(\frac{\log{r}}{r^5}\right),\\
T_7\longrightarrow &-\frac{1}{4r^2}+\frac{i\omega}{6r^3}+\frac{t_7}{r^4}-\frac{1}{16r^4} (\omega^2-q^2)\log{r} + \mathcal{O}\left(\frac{\log{r}}{r^5}\right),\\
T_8\longrightarrow &\frac{t_8}{r^4}+\frac{\log{r}}{24r^4} + \mathcal{O} \left(\frac{\log{r}}{r^5}\right),\\
T_9\longrightarrow &-\frac{i\omega}{r}-\frac{1}{4r^2}(q^2+\omega^2)+\frac{1}{12r^3}i \omega(3q^2-\omega^2)+\frac{t_9}{r^4}\\
&+\frac{1}{16r^4}(q^2-\omega^2)^2\log{r} + \mathcal{O}\left(\frac{\log{r}}{r^5}\right).
\end{split}
\end{equation}
Then, from equation~(\ref{k eq}), we can obtain the large $r$ limit of $k$
\begin{equation}\label{asym k}
\begin{split}
k\xlongrightarrow{r\to\infty}~&\frac{2}{3}\left(r^3+i\omega r^2\right)\partial u-\frac{2}{3}r^3\partial_k h_{0k}+\frac{1}{3}r^2\partial^2h_{00}-\frac{1}{6}r^2 \left(\partial_i\partial_j h_{ij}- \partial^2h_{kk}\right)\\
&+\frac{1}{36}\left[-4i\omega q^2 \partial_k h_{0k} -2q^2 \partial^2h_{00}+ (\omega^2-q^2) \partial^2h_{kk}\right.\\
&\left.+(q^2-3\omega^2)\partial_k \partial_l h_{kl} \right]\log{r}- \frac{5}{216}\left[ 4i\omega q^2 \partial_k h_{0k} + 2q^2 \partial^2 h_{00} +\right.\\
&\left.\left(q^2-\omega^2\right) \partial^2 h_{kk} + (3\omega^2-q^2) \partial_k \partial_l h_{kl}\right]+\mathcal{O} \left(\frac{1}{r}\right).
\end{split}
\end{equation}
The integration constants in $V_i$'s and $k$ are fixed by the convention $\Pi_{\langle0i\rangle}=0$ and $\Pi_{\langle 00\rangle}=0$, respectively. The asymptotic requirements~(\ref{AdS constraint}) were used to determine the leading behaviors in the near-boundary expansion. Our problem is now reduced to finding  a set of pre-asymptotic coefficient functions $t_i$, which propagate into final expression for the boundary
stress tensor (eq.~(\ref{emt1})).  These coefficients can be computed only through full integration from the horizon to the boundary. The regularity conditions at the horizon ($r=1$) provide sufficient  initial data to uniquely fix all the $t_i$'s, and consequently all the transport coefficient functions introduced in~(\ref{diss}).

We also notice the logarithmic branches in~(\ref{asym 34},\ref{asym 56789},\ref{asym k}), which emerge solely due to the boundary metric perturbation $h_{\mu\nu}$, as is clear from~(\ref{decomposition}). The logarithms appear starting from the fourth order in the boundary metric derivatives, in full agreement with computations of~\cite{hep-th/9806087,hep-th/0010138,hep-th/0002230}.

The stress-energy tensor for the boundary CFT is obtained by substituting the above large $r$ behaviors into~(\ref{stress tensor1},\ref{stress tensor2}),
\begin{equation}\label{emt1}
\left\{
\begin{aligned}
T_{00}=&\,3-12\epsilon {\bf b}_1-3\epsilon h_{00},\\
T_{0i}=&\,T_{i0}=-4\epsilon u_i+\epsilon h_{0i},\\
T_{ij}=&\,\delta_{ij}(1-4\epsilon {\bf b}_1)+\epsilon h_{ij}+\epsilon \left\{ 8t_1 \sigma_{ij}+4t_2 \pi_{ij}\right. \\
&\left.+\left[8(t_1+t_3)+\frac{1}{24}i\omega(3q^2-7\omega^2)\right] \tilde{\sigma}_{ij}+ \left[4(t_2+t_4)+\frac{5}{36}i\omega\right]\tilde{\pi}_{ij}\right.\\
&\left.+\left[4t_5-\frac{1}{144}(q^2+21\omega^2)\right]\varphi_{ij}+\left[2i\omega t_2+ 4t_6+\frac{1}{144} (13q^2-\omega^2)\right]\chi_{ij}\right.\\
&\left.+\left[8t_7-\frac{1}{8}(\omega^2+3q^2)\right]\psi_{ij}+\left(4t_8-\frac{7}{72}\right) \phi_{ij}\right.\\
&\left.+\left[4i\omega t_1+4t_9+\frac{1}{48}(7\omega^4-9q^4-6\omega^2q^2)\right]\tau_{ij} \right\}.
\end{aligned}
\right.
\end{equation}
The tensors $\sigma_{ij}$ and $\pi_{ij}$ are covariant versions of those defined in~\cite{1406.7222,1409.3095},
\begin{equation}
\begin{split}
\sigma_{ij}&\equiv\frac{1}{2}\left(\nabla_iu_{j}+\nabla_{j}u_i-\frac{2}{3}g_{ij} \vec{\nabla}\vec{u}\right)=\hat{\sigma}_{ij}-\tilde{\sigma}_{ij}-\frac{1}{2}i\omega \tau_{ij},\\
\pi_{ij}&\equiv\nabla_{i}\nabla_{j}\nabla u-\frac{1}{3}g_{ij} \vec{\nabla}^2\nabla u=\hat{\pi}_{ij}-\tilde{\pi}_{ij}-\frac{1}{2}i\omega \chi_{ij},
\end{split}
\end{equation}
where the linearization approximation~(\ref{linearization}) was employed in expressing them in terms of the basic structures defined in Table~\ref{table1}. Combining non-derivative parts in~(\ref{emt1}), we arrive at the standard expression for conformal ideal fluid,
\begin{equation}
T_{\mu\nu}^{\textrm{Ideal}}=\frac{1}{{\bf{b}}^4}\left(g_{\mu\nu}+4u_{\mu}u_{\nu}\right).
\end{equation}
The expansion coefficients $t_i$'s are not fully independent. There are four constraints among them, derivation of which is deposited to~Appendix~\ref{appendix3}:
\begin{equation}\label{constraint ti}
\begin{split}
0=&\,t_9-i\omega t_3-q^2t_7,\\
0=&\,2(t_1+t_3)-2i\omega t_5-q^2(t_2+t_4),\\
0=&\,2(t_6+t_7)-i\omega t_4-2q^2 t_8,\\
0=&\,(t_5-t_6)-i\omega t_4-q^2t_8.
\end{split}
\end{equation}
Eq.~(\ref{emt1}) can be rewritten in terms of the Weyl tensor $C_{\mu\alpha\nu\beta}$ and its derivatives. With the help of Appendix~\ref{appendix1} and after some algebraic manipulations including the constraints~(\ref{constraint ti}), the dissipation tensor $\Pi_{\mu\nu}$ takes the form~(\ref{diss}) with the transport coefficients $\eta$ etc being expressed as combinations of the coefficients $t_i$'s,
\begin{equation}\label{relation}
\begin{split}
\eta&=-4t_1,~~~~~\zeta=-4t_2,~~~~~\theta=-6t_2\\
\kappa&=\frac{1}{6}\left[-48(t_5+t_6)+24i\omega (t_2+2t_4)+\omega^2(48 t_8-1) -q^2 \right],\\
\rho&=-\frac{4}{3}\left[12(t_2+t_4)+i\omega(1-24 t_8)\right],\\
\xi&=\frac{1}{12}\left(7-288t_8\right).
\end{split}
\end{equation}
Notice that, in addition to the explicit $\omega$ and $q$ dependencies present in~(\ref{relation}), all the coefficient functions $t_i$ also depend on these momenta.

As has been emphasized above, the stress-energy tensor~(\ref{emt1}) is off-shell: so far we have managed to establish a map between the dynamical components of~(\ref{einstein eq}) and the transport coefficients. In our procedure, we have specified neither the velocity nor the boundary metric perturbation. We expect that the conservation laws $\nabla^{\mu}T_{\mu\nu}=0$ should emerge from the constraints in~(\ref{einstein eq}), more precisely, from the constraints $E_{vv}=E_{vi}=0$. We have checked that it is indeed the case. The relevant computations are deferred to Appendix~\ref{appendix4}.

\subsection{Gravitational susceptibilities of the fluid} \label{subsection43}

We have now reached the point where we are  to solve the bulk equations~(\ref{v12t12}-\ref{V8}). These differential equations have to be evolved from the horizon to the conformal boundary with appropriate boundary conditions imposed at both ends. For the sector~(\ref{v12t12}) this calculation has been performed in~\cite{1409.3095} and revealed the viscosity functions. Here we apply the methods developed in~\cite{1409.3095} to other sectors. We first perform analytical analysis in the hydrodynamic limit and then full numerical study for generic values of momenta.
Below we merely summarize the results, while referring the reader to~\cite{1409.3095} for calculational details.

\subsubsection{Analytical results: hydrodynamic expansion}\label{subsubsection431}
We start with the hydrodynamic limit $\omega,~q_i\ll 1$ and solve the equations~(\ref{v12t12}-\ref{V8}) perturbatively. We introduce a formal parameter $\lambda$ by $\omega\to \lambda \omega,~q_i\to \lambda q_i$, and expand $V_i$'s and $T_i$'s in powers of $\lambda$
\begin{equation}\label{VT exp}
V_i(\omega,q_i,r)=\sum_{\lambda=0}^{\infty}\lambda^n V_i^{(n)}(\omega,q_i,r), ~~~~~~~ T_i(\omega,q_i,r)=\sum_{\lambda=0}^{\infty}\lambda^n T_i^{(n)}(\omega,q_i,r),
\end{equation}
where $\lambda$ will be eventually set to unity. At every order in $\lambda$, a system of ordinary second order differential equations can be integrated~\cite{1409.3095}. The expansion coefficients $V_i^{(n)}$ and $T_i^{(n)}$ are expressible as double integrals, which are summarized in Appendix~\ref{appendix5}. Their $r\rightarrow \infty$ asymptotic behavior reveals the coefficients $t_i$'s:
\begin{eqnarray}
\begin{split}
&t_1=-\frac{1}{4}+\frac{\log{2}-2}{8}i\omega+\frac{1}{192}\left\{6q^2+\omega^2\left[6 \pi-\pi^2+12(2-3\log{2}+\log^2{2})\right]\right\}+\cdots,\\
&t_2=-\frac{1}{48}\left(5-\pi-2\log{2}\right)+\cdots,~~~~t_3=\frac{1}{4}- \frac{\log{2}-1}{8}i\omega+\cdots,\\
&t_4=-\frac{1}{48}\left(1+\pi+2\log{2}\right)+\cdots,~~~t_5=-\frac{1}{8}- \frac{1}{32} \left(4+\pi-2\log{2}\right)i\omega+\cdots,\\
&t_6=-\frac{1}{8}-\frac{1}{96}\left(10+\pi-10\log{2}\right)i\omega+\cdots,~~~t_7= \frac{1}{8}-\frac{1}{32}\left(4\log{2}-3\right)i\omega+\cdots,\\
&t_8=\frac{13-12\log{2}}{288} +\cdots,~~~t_9=\frac{1}{4}i \omega+ \frac{1}{8} \left[q^2+(\log{2}-1)\omega^2\right]+\cdots.
\end{split}
\end{eqnarray}
When substituted in (\ref{relation}), this leads to the expansions~(\ref{emt expansion}) for the viscosities and GSFs.
\subsubsection{Numerical results: all-order resummed hydrodynamics}\label{subsubsection432}
We are now to solve the equations~(\ref{v12t12}-\ref{V8}) for generic values of $\omega$ and $q^2$. Since the boundary conditions are imposed at different points, we resort to the \emph{shooting technique} and solve them numerically. As in~\cite{1409.3095}, we first start with a trial solution (initial condition) at one boundary (horizon) and integrate these  equations to the second boundary (the conformal boundary). Thus generated solution should match the boundary conditions at the end of the integration. So, the trial solution has to be finely tuned in order to have the boundary conditions at the end of the integration satisfied. This fine-tuning procedure is mapped into an optimization problem.
\begin{figure}[htbp]
\centering
\includegraphics[scale=0.55]{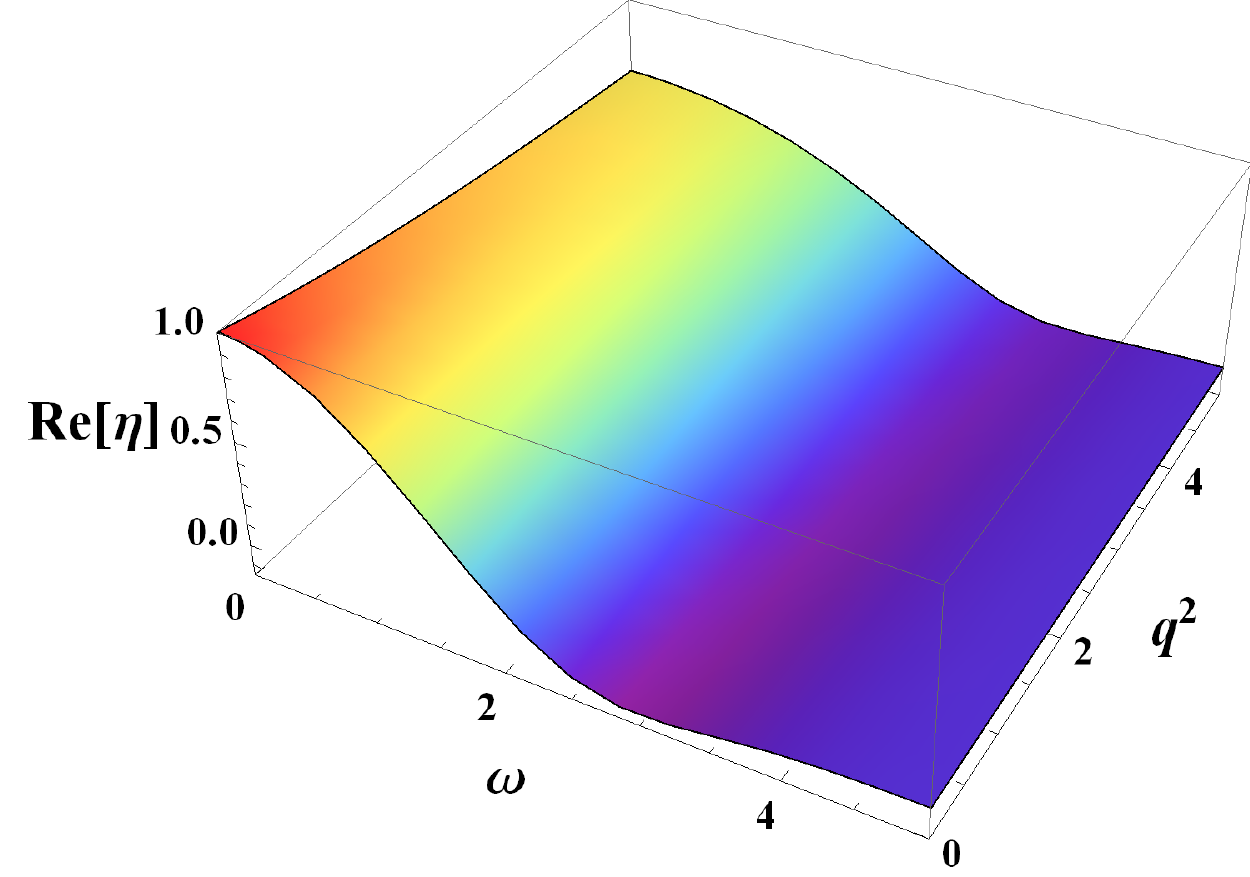}
\includegraphics[scale=0.55]{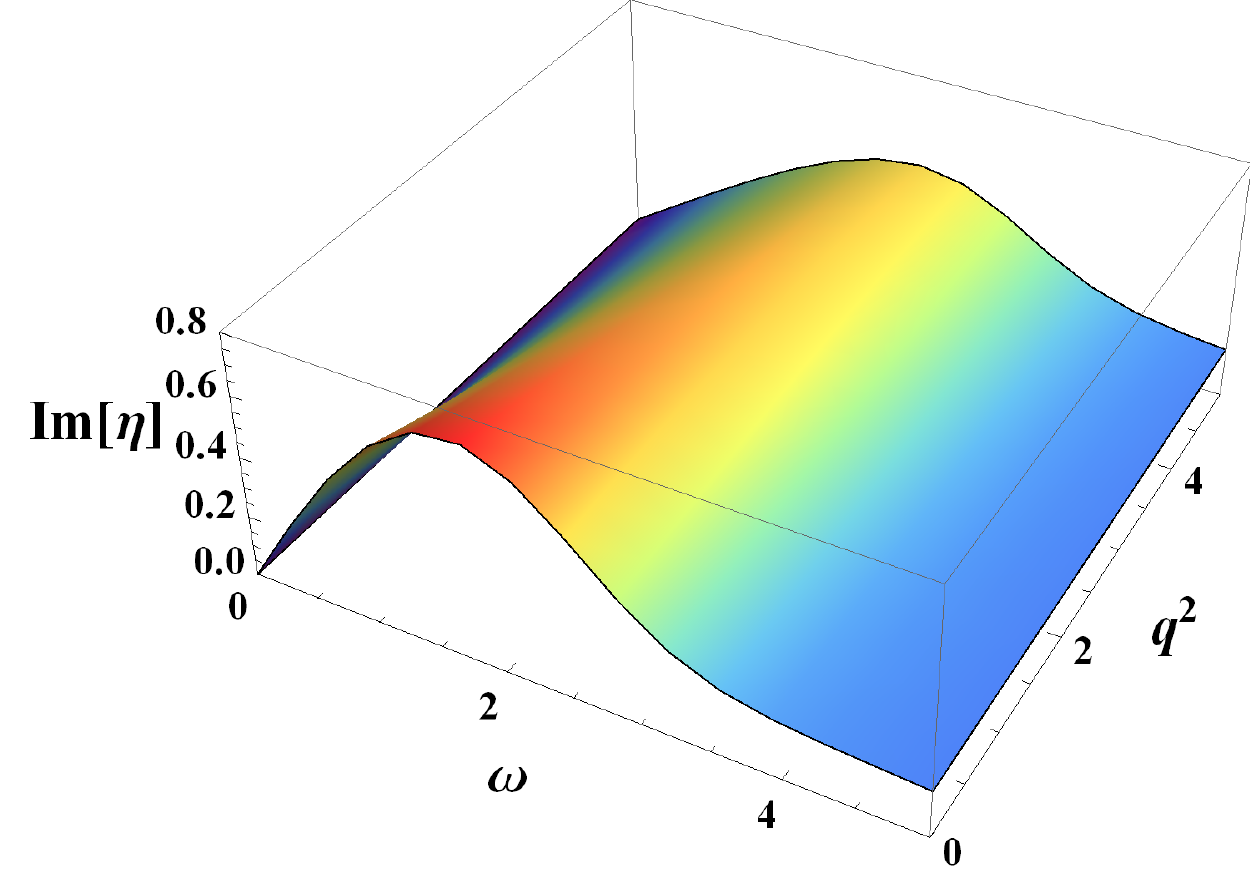}
\caption{Viscosity function $\eta$ as a function of $\omega$ and $q^2$.}
\label{figure1}
\end{figure}
\begin{figure}[htbp]
\centering
\includegraphics[scale=0.55]{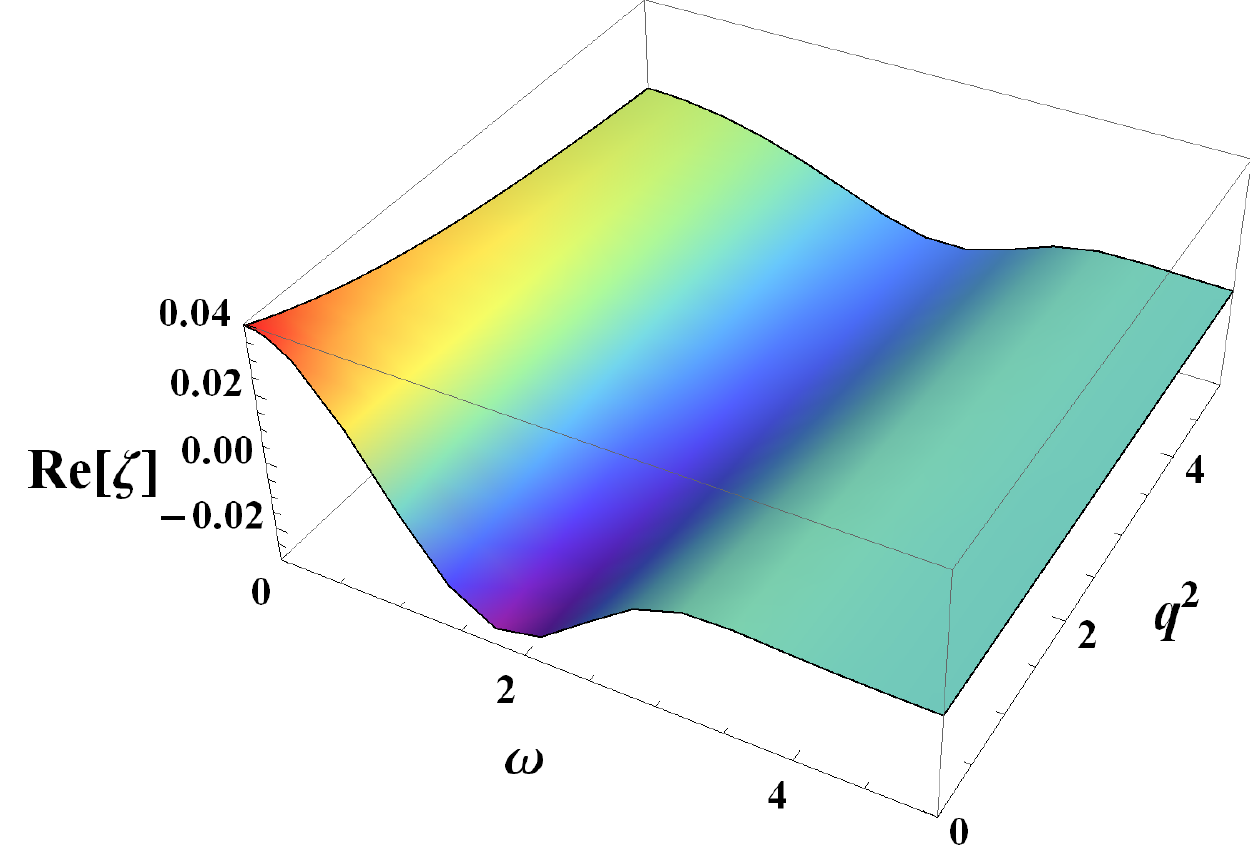}
\includegraphics[scale=0.55]{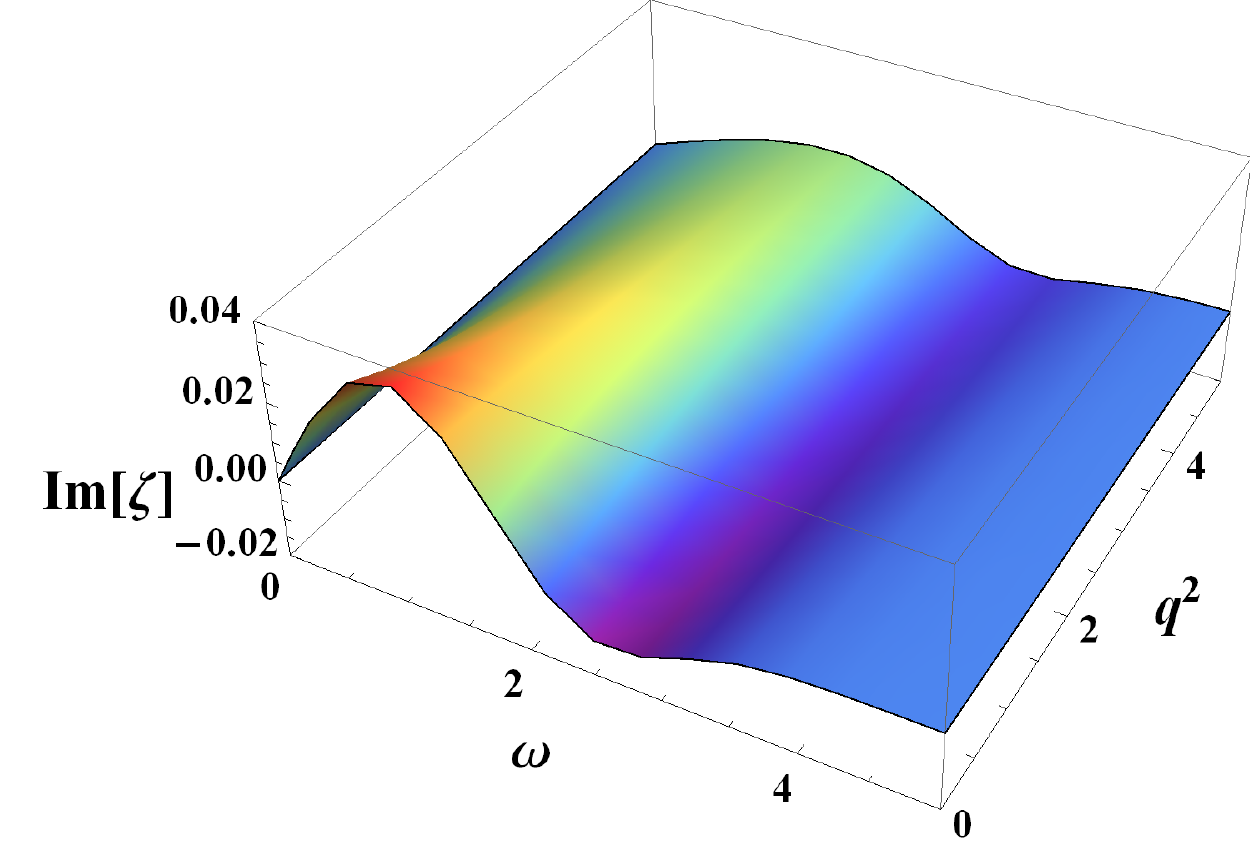}
\caption{Viscosity function $\zeta$ as a function of $\omega$ and $q^2$.}
\label{figure2}
\end{figure}
\begin{figure}[htbp]
\centering
\includegraphics[scale=0.52]{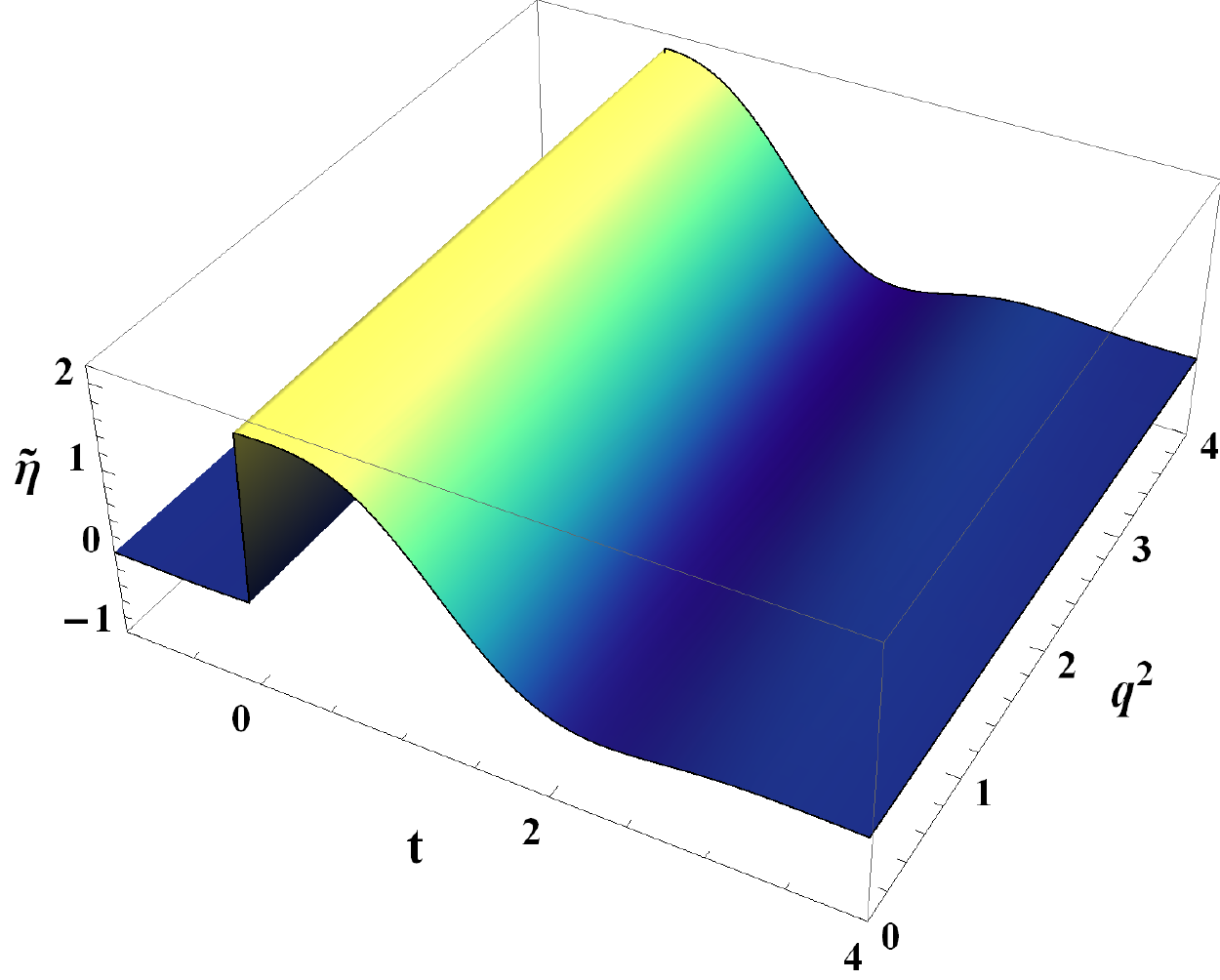}
\includegraphics[scale=0.85]{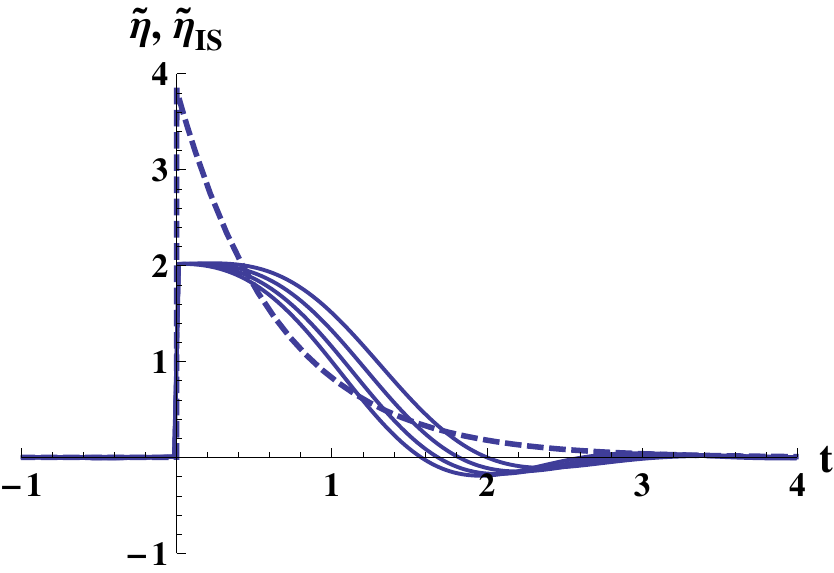}
\caption{Memory function $\tilde{\eta}\left(t,q^2\right)$. Left: 3D plot as a function of time $t$ and $q^2$. Right: 2D plots as functions of time $t$; the dashed curve corresponds to Israel-Stewart hydrodynamics with $\eta_0=1$ and the relaxation time $\tau_{R}=\left(2-\log{2}\right)/2$; the solid curves display our result
at  different $q^2$ (from the rightmost, $q^2=0,1,2,3$).}
\label{figure3}
\end{figure}
\begin{figure}[htbp]
\centering
\includegraphics[scale=0.55]{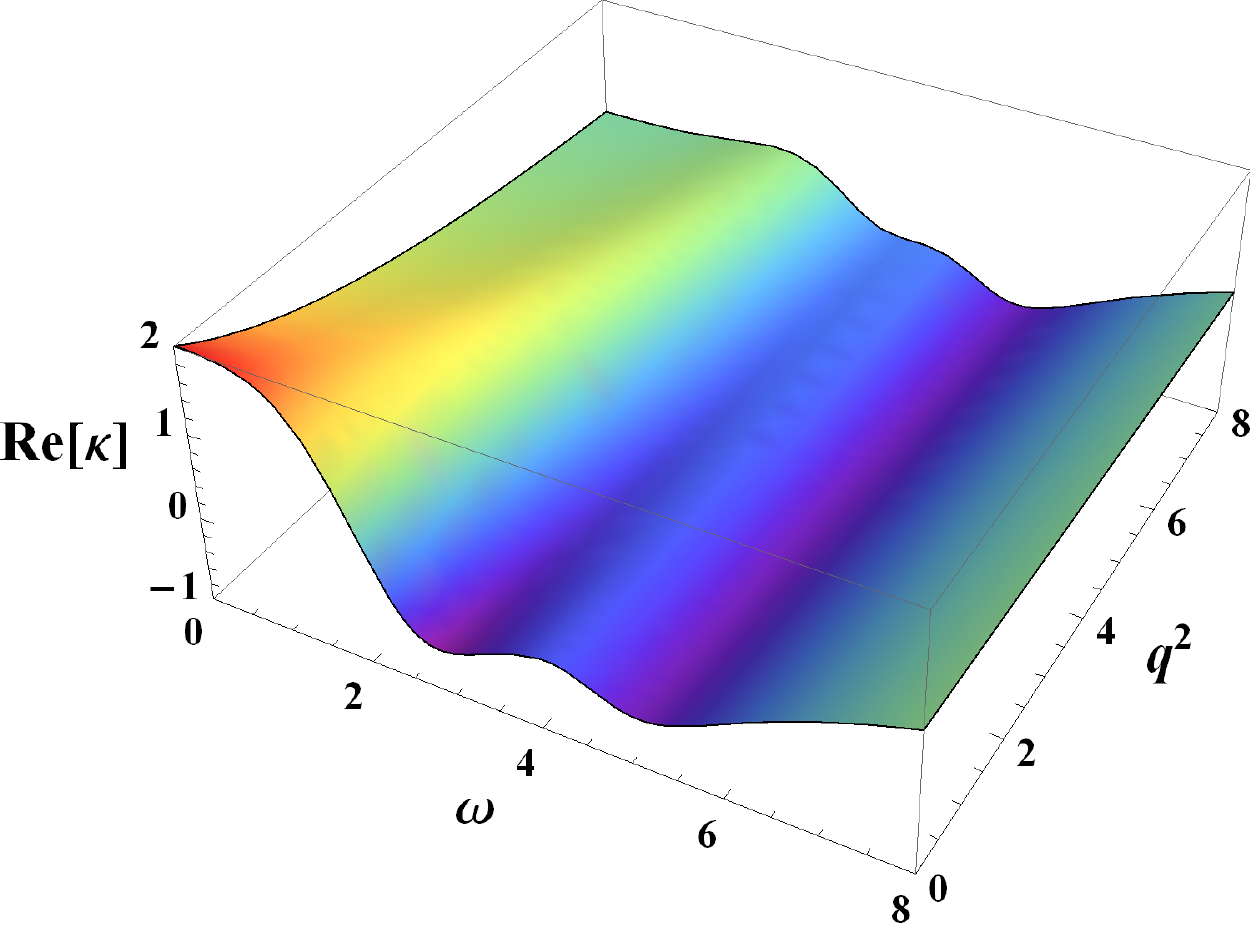}
\includegraphics[scale=0.55]{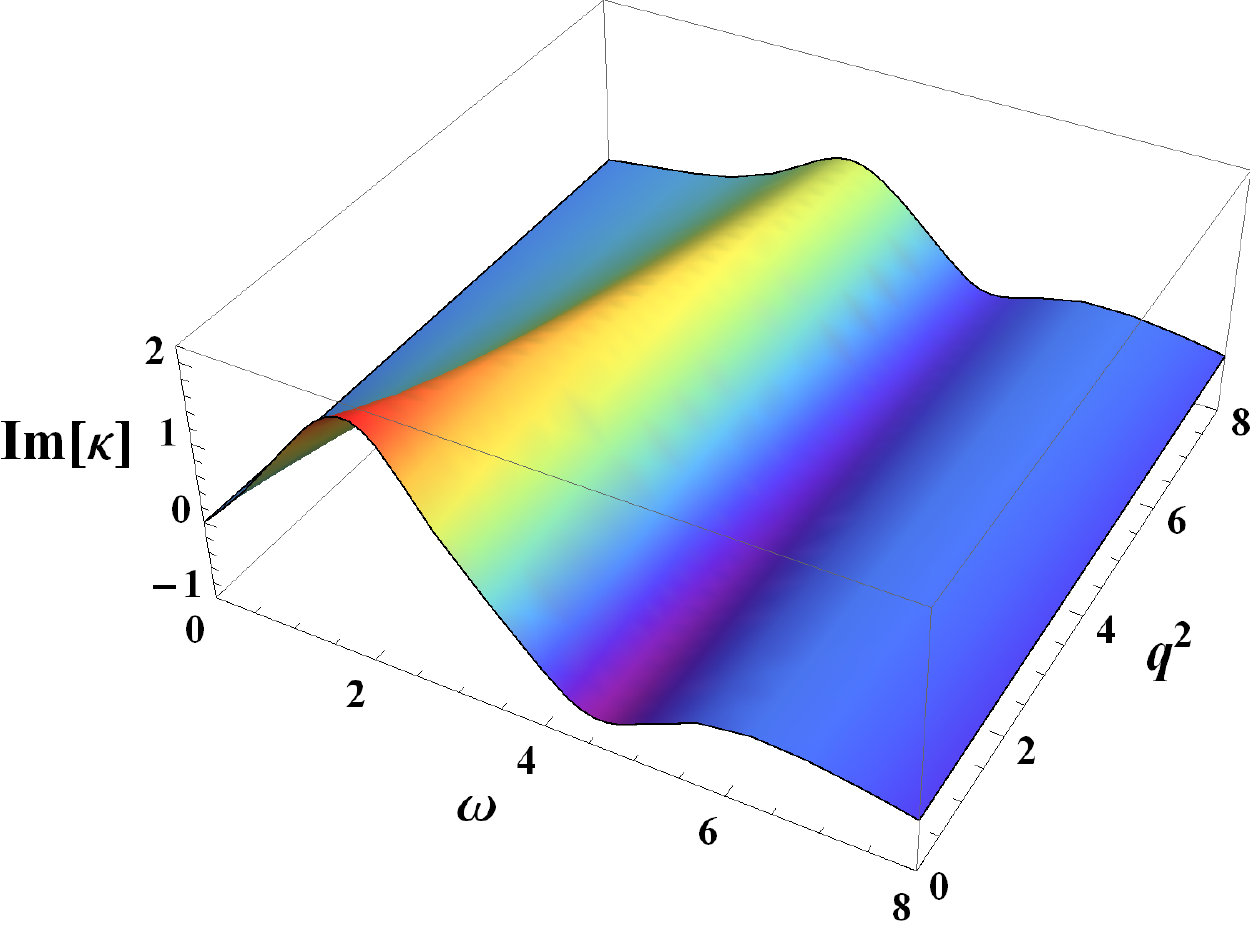}
\caption{GSF $\kappa$ as function of $\omega$ and $q^2$.}
\label{figure4}
\end{figure}
\begin{figure}[htbp]
\centering
\includegraphics[scale=0.55]{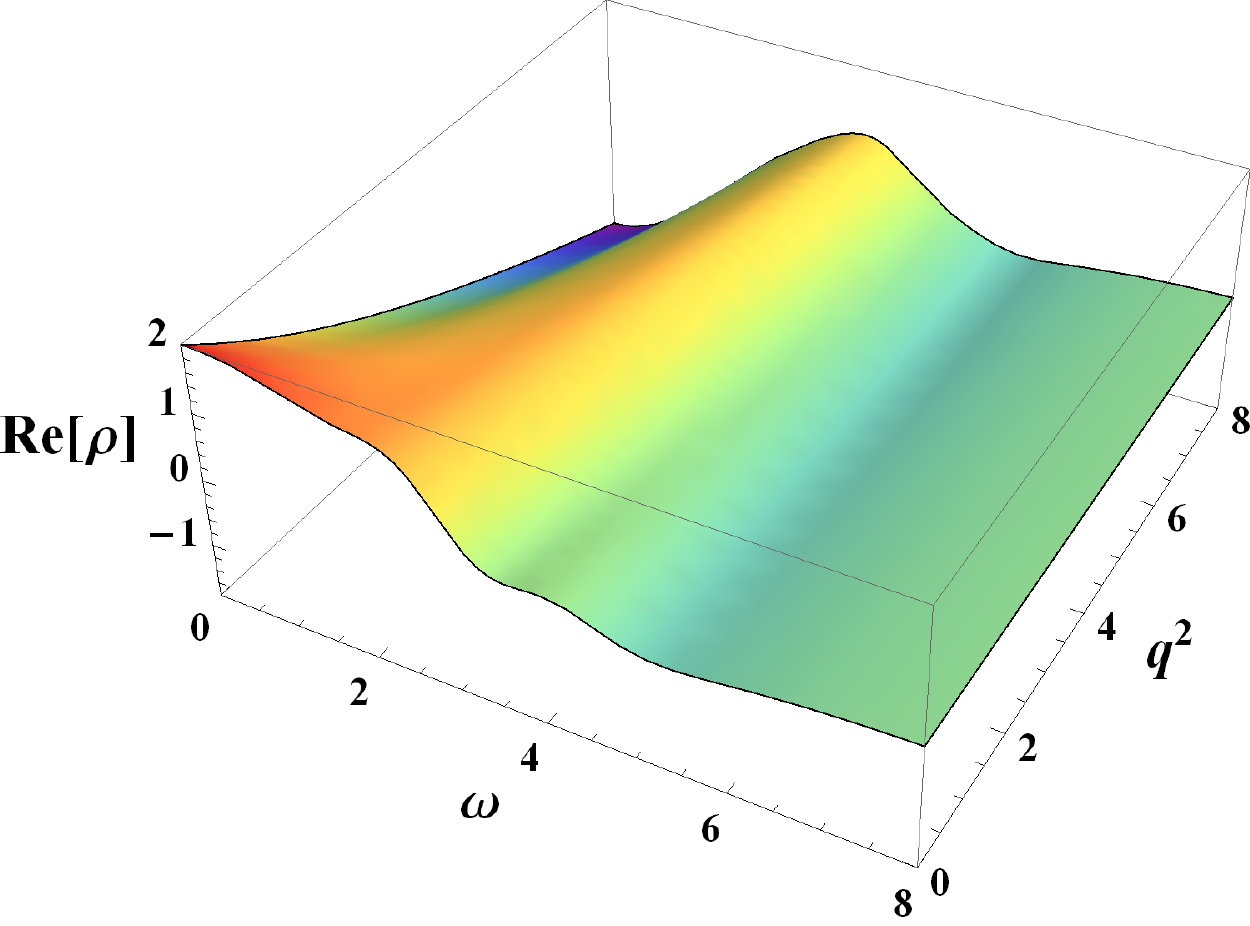}
\includegraphics[scale=0.55]{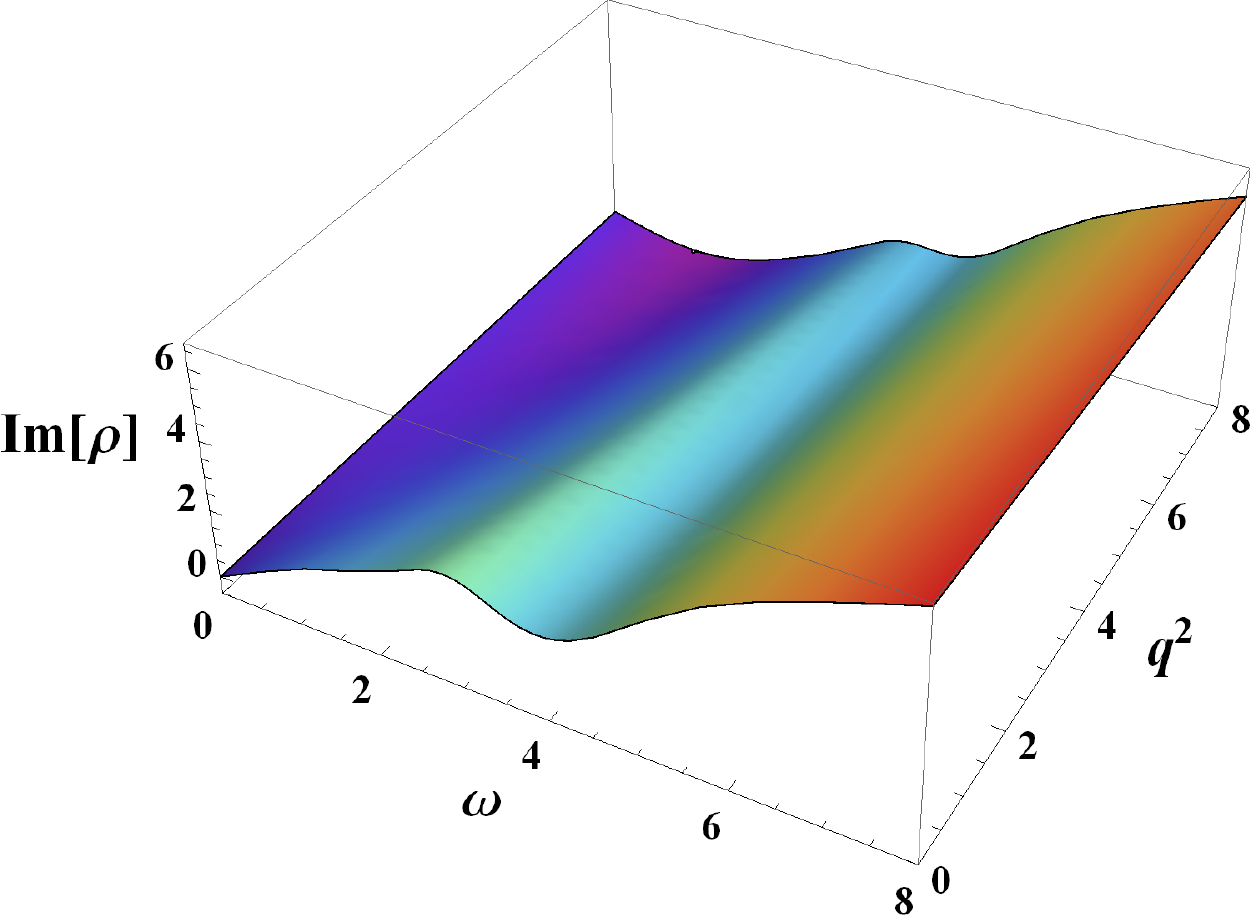}
\caption{GSF $\rho$ as function of $\omega$ and $q^2$.}
\label{figure5}
\end{figure}
\begin{figure}[htbp]
\centering
\includegraphics[scale=0.55]{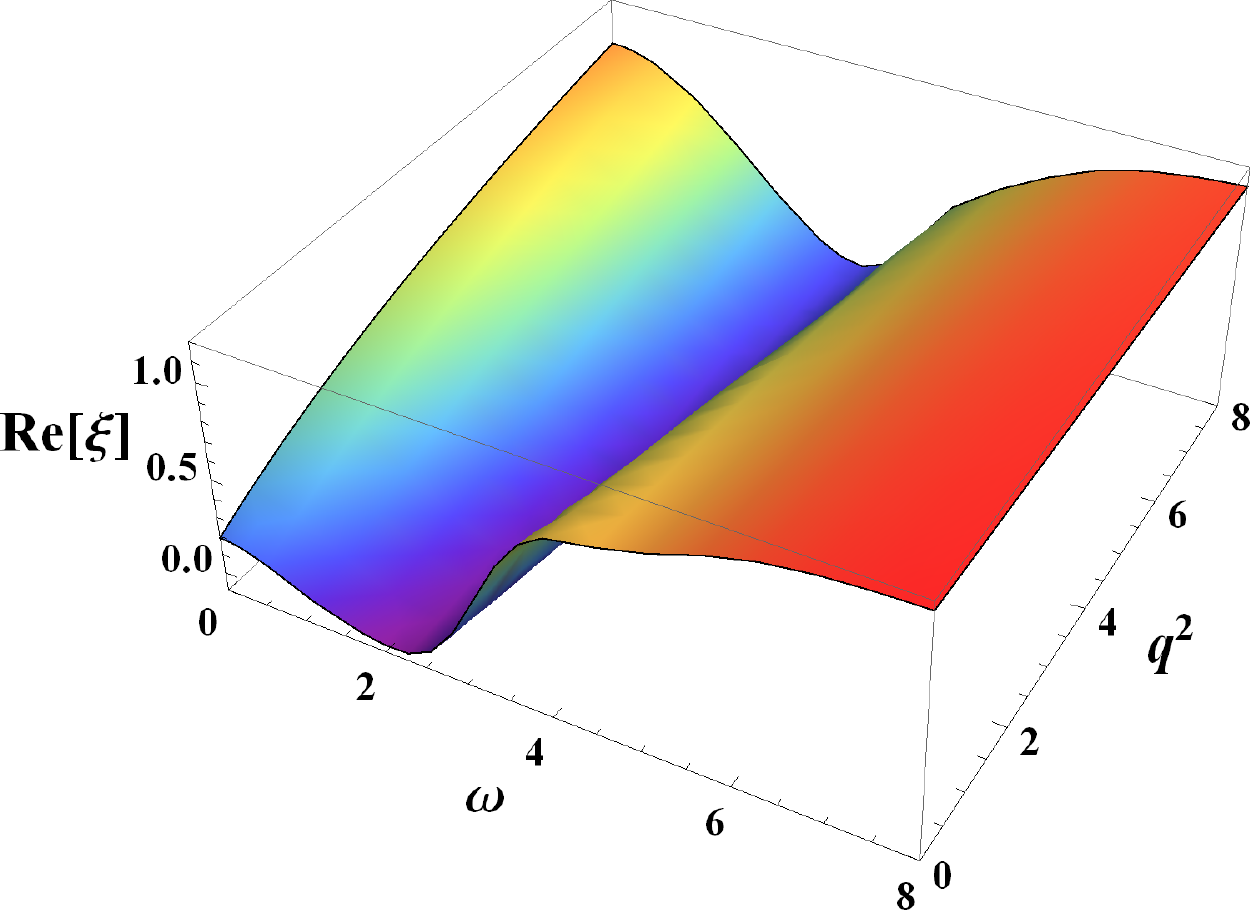}
\includegraphics[scale=0.55]{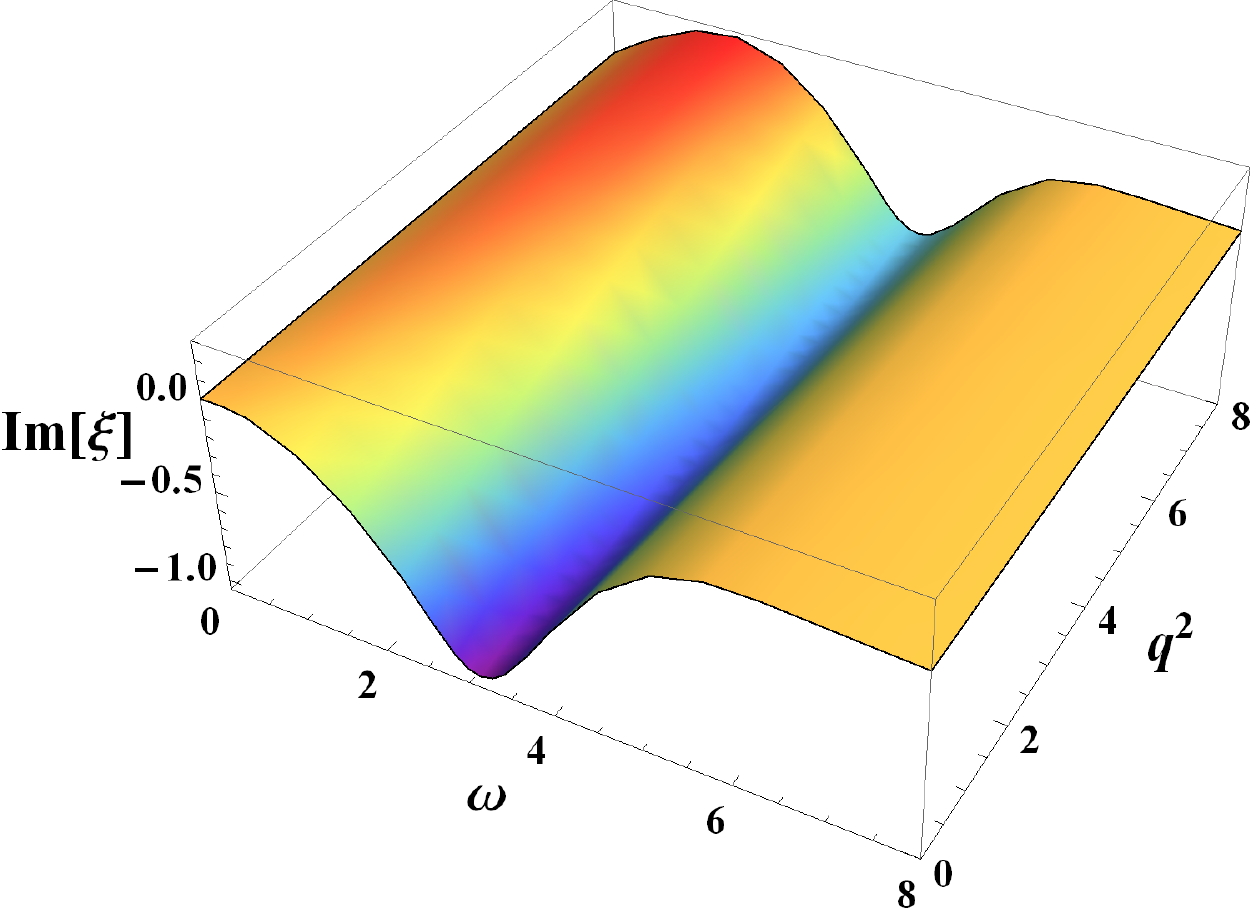}
\caption{GSF $\xi$ as function of $\omega$ and $q^2$.}
\label{figure6}
\end{figure}

The viscosities $\eta$ and $\zeta$, originally computed in~\cite{1409.3095}, are shown in figures~\ref{figure1} and~\ref{figure2}. $\eta$ and $\zeta$ are observed to vanish at very large momenta. As has been pointed out in section~\ref{section2}, it is worth looking at the corresponding memory function defined through the inverse Fourier transformation,
\begin{equation}
\tilde{\eta}\left(t,q^2\right)=\frac{1}{\sqrt{2\pi}}\int_{-\infty}^{\infty} \eta\left(\omega,q^2\right)e^{-i\omega t}d\omega.
\end{equation}
The viscosity $\eta$ is weakly dependent on spatial momenta, meaning it can be thought of as almost a local function of space coordinates. For this reason, we have chosen not to Fourier transform in $q$ and present the memory function in a mixed $(t,q^2)$ representation. In figure~\ref{figure3}, we show the time dependence of the memory function $\tilde{\eta}(t,q^2)$ and compare it with the Israel-Stewart hydrodynamics, for which the viscosity function $\eta_{IS}(\omega)=\eta_0/(1-i\omega \tau_{R})$ and the memory function is
\begin{equation}
\tilde{\eta}_{IS}(t)=\frac{\sqrt{2\pi}}{\tau_{R}}\eta_0 e^{-t/\tau_{R}}\Theta(t).
\end{equation}
Notice that the memory function $\tilde{\eta}$ vanishes exactly for negative times, in accordance with the causality requirement, as discussed in section~\ref{section2}. For positive times, the memory function decreases with time, followed by damped oscillations, and eventually approaches zero at very late times.

We are now to present the results for the GSFs. As has been proven above, $\theta=3\zeta/2$. Numerical results for the remaining GSFs are displayed in figures~\ref{figure4},~\ref{figure5} and~\ref{figure6}. Compared to the viscosities,
we observe a noticeably different behavior of the GSFs as functions of momenta. Particularly, some components do not vanish at large momenta. Below we shall provide interpretation of such behavior.

$\kappa$ is found to depend weakly on spatial momentum $q_i$, somewhat similarly to the viscosities $\eta$ and $\zeta$. Higher order derivative terms reduce $\textrm{Re}\left[\kappa\right]$ until the point $\left\{\omega\approx 2.6,q^2=0\right\}$, where $\textrm{Re}[\kappa]$ reaches its minimum. At larger frequencies, $\textrm{Re}\left[\kappa\right]$ oscillates around zero and then it starts to grow around $\omega\sim 5$. The imaginary part of $\kappa$ increases from zero to its maximum at $\left\{\omega\approx 1.8,q^2=0\right\}$. Then, $\textrm{Im}[\kappa]$ decays very fast until $\left\{\omega\approx 4.5,q=0\right\}$ where it starts to oscillate around zero. At very large momenta, the real part of $\kappa$ keeps growing while its imaginary part vanishes.

The GSFs $\rho$ and $\xi$ behave rather differently from their cousin $\kappa$. When $\omega\lesssim2$, $\textrm{Re}[\rho]$ drops sharply as the spatial momentum is increased. For larger $\omega$, the real part of $\rho$ first decays rather slowly until the point $\left\{\omega\approx2.0,q^2=0\right\}$, where $\textrm{Re}\left[\rho\right]$ starts to drop quickly and eventually approaches zero at very large momenta. In contrast, the imaginary part of $\rho$ grows fast at large frequencies. The spatial derivatives also affect $\textrm{Re}[\xi]$, in the same region of $\omega\lesssim2$. With $\omega$ increased, $\textrm{Re}[\xi]$ grows fast till $\omega\approx4$, where it tends to approach a constant value of $1.2$. The imaginary part of $\xi$ weakly depends on spatial momentum. It first decreases from zero till certain minimum around $\left\{\omega\approx3.2,q=0\right\}$, where it starts to increase as function of $\omega$ but approaches zero starting from $\omega\approx5$.

We do not have a good intuitive explanation of the observed behaviors of the GSFs from small to mid momenta $\omega\sim q\sim 5$. Nevertheless, the large momentum asymptotic region can be understood. In the limit $\omega,q\rightarrow\infty$ we expect to recover
zero temperature limit of the theory. The viscosity functions vanish in this limit, whereas the GSFs do not. While the role of each individual GSF coefficient is not obvious, their combinations enter as residues in the expression for two point correlators~\cite{0905.4069}. We review the correlators in Appendix~\ref{appendix4}.  Thus we expect the GSFs to play a role in recovering the zero temperature limit of the correlators, which are known analytically and also quoted in Appendix~\ref{appendix4}. We have checked numerically that, starting from $\omega\sim q\sim 5$, the correlators computed using the GSFs (eqs.~(\ref{scalar channel},\ref{shear channel},\ref{sound channel})) reproduce well their zero temperature counterparts~(\ref{zero}).

From the zero temperature limit of the correlators, we can analytically deduce the asymptotic behaviors of the GSFs,
\begin{equation}
\begin{split}
\kappa\sim &\left(\omega^2-q^2\right)^2 \log{\left(\omega^2-q^2 \right)},\\
\rho\sim &\frac{i}{\omega}\left(\omega^2-q^2\right)\log{\left(\omega^2-q^2\right)},\\
\xi\sim &\log{\left(\omega^2-q^2\right)}
\end{split}
\end{equation}
which agrees with the numerical results presented above.

\section{Summary and discussion}\label{section5}

In this work we introduced  (non-dynamical) metric perturbations of the flat boundary, and following the the setup of~\cite{1406.7222,1409.3095} obtained an asymptotically locally $\textrm{AdS}_5$ black brane solution to the Einstein gravity. Our construction is limited to linear order in amplitudes of the fluid velocity and boundary metric perturbation, but exact in terms of the derivative expansion. This gravity solution is holographically dual to a relativistic conformal hydrodynamics in a weakly curved background spacetime. This dual hydrodynamics is equivalent to full linear response theory with the background metric perturbation playing the role of external disturbance. Since all order derivatives are included in the stress tensor, the effective theory constructed here is causal and expected to be free of any instabilities.

The constitutive relation~(\ref{diss}) for the stress-energy tensor emerged from the dynamical components of the bulk Einstein equations, with holographically determined values of transport coefficients. Overall we introduced six transport coefficient functions: two viscosities and four GSFs, transport coefficient functions associated with fluid's response to curvature perturbations. These terms vanish in flat space and do not affect hydrodynamic equations. When a non-zero gravitational perturbation $h_{\mu\nu}$ is turned on, one can consider it as an external force such that the dual fluid becomes  forced in Minkowski space~\cite{0806.0006}. The arguments of~\cite{0806.0006} indicate that different gravitational forces are able to stir the fluid into flows with velocity differences of order one.

It appears that the GSFs have a larger role beyond being naive linear response coefficients. As was first noticed in \cite{0712.2451} and further elaborated in~\cite{0905.4069}, GSFs enter the expressions for thermal correlators of $T_{\mu\nu}$.
In Appendix~\ref{appendix4}, the correlators are derived using the linear response theory from the constitutive relation~(\ref{diss}). The shear and sound wave modes predicted from hydrodynamic equations in flat spacetime appear as poles in these  correlators. Residues of the thermal correlators get contributions from both the viscosity $\eta$ and all the GSFs (see~(\ref{scalar channel},\ref{shear channel},\ref{sound channel})).
We have found that some of the GSFs do not vanish at large momenta. This is consistent with the expectations put forward in~\cite{0905.4069}. It was argued there that zero temperature expressions for the correlators have to be recovered from the thermal ones in the large momenta limit and that is what the GSFs take care of. The thermal correlators incorporate information about both thermal physics related to matter flow and the vacuum contribution due to pair production. Furthermore, there are interference effects between these two types of physics, as encoded in various GSF terms.

A remark about Weyl symmetry is in order. The stress tensor of conformal fluid is supposed to transform covariantly under the Weyl rescaling ${g}_{\mu\nu}\rightarrow e^{-2\phi} g_{\mu\nu}$~\cite{0806.0006,0809.4272}. In~\cite{0712.2451}, the symmetry was used to impose selection rules on the form of second order relativistic conformal fluid. It was subsequently exploited in~\cite{0905.4069} in order to constrain the general form of linearly resummed constitutive relation\footnote{Thanks to linearization, the Weyl anomaly is irrelevant for the discussion.}. Both the Navier-Stokes term $\nabla_{\mu}u_{\nu}$  and the Weyl tensor $C_{\mu\alpha\nu\beta}$ transform as needed under the Weyl transformation. It is, however, obvious that applying extra derivatives on them breaks the symmetry. It turns out that the linearization does not commutate with the Weyl transformation: the nonlinear terms like $\mathcal{A}_{\lambda}Q^{\mu\cdots}_{\nu\cdots}$ (see Appendix~\ref{appendixw} for introduction to Weyl-covariant formalism) do generate some linear pieces in the Weyl transformed spacetime $\tilde{g}_{\mu\nu}=e^{-2\phi} g_{\mu\nu}$. The Weyl-covariance of the fluid stress-energy tensor is to be recovered via a non-linear completion of the expression~(\ref{diss})~\cite{0905.4069}. In order to have a manifestly Weyl-covariant form of the stress-tensor, we can substitute the normal derivative $\nabla_{\mu}$ by
Weyl-covariant long-derivative $\mathcal{D}_{\mu}$ introduced in~\cite{0801.3701} (see Appendix~\ref{appendixw}). With this procedure we would also recover a significant fraction of the non-linear terms neglected by our analysis.

We have been emphasizing the relation between all-order gradient resummation and  causality. Particularly, in section~\ref{section2} we argued that causality induces constraints on positions of the poles of the transport coefficients. We demonstrated this effect by explicitly computing the viscosity memory function, which indeed does not have any support in future. In a forthcoming publication~\cite{BLS}, we study all order linearized fluid dynamics whose dual gravity involves a Gauss-Bonnet correction. We find that the value of the Gauss-Bonnet coupling can be constrained by  causality requirement on the linearly resummed fluid dynamics.

We hope our results may share some generic features with those of realistic QCD and could be helpful in understanding the nature of hydrodynamic gradient expansion. In fact, when resumming all order derivative terms, we explicitly go beyond the conventional hydrodynamic limit and account for non-hydrodynamic modes in our construction. At momenta of order one the hydrodynamic and non-hydrodynamic modes mix and only full resummation
makes sense.

The viscosity memory function computed here could be useful for phenomenological applications. We have illustrated how different it is from the naive second order ansatz  a la Israel-Stewart. In \cite{0704.1647}, higher order gradients  were argued to have large phenomenological effects. So, it is quite important to better explore these effects by simulating hydro evolution of quark-gluon plasma with our viscosity memory function, despite its obviously limited applicability to real QCD.

\appendix
\section{Boundary curvatures}\label{appendix1}
In this appendix, we summarize expressions for  boundary curvatures that are useful in constructing $T_{\mu\nu}$. We work with the mostly plus signature for the metric tensors. Thus, the Riemann curvature of $g_{\mu\nu}$ is conventionally defined as
\begin{equation}
R^{\lambda}_{~\mu\sigma\nu}=\partial_{\sigma}\Gamma_{\mu\nu}^{\lambda}-\partial_{\nu} \Gamma_{\mu\sigma}^{\lambda}+\Gamma^{\kappa}_{\mu\nu}\Gamma_{\kappa\sigma}^{\lambda} - \Gamma^{\kappa}_{\mu\sigma}\Gamma^{\lambda}_{\kappa\nu},
\end{equation}
where $\Gamma_{\mu\nu}^{\lambda}$ is the Christoffel symbol. To linear order in $h_{\mu\nu}$, $R_{\mu\nu}$ and $R$ have the expressions,
\begin{equation}
\begin{split}
R_{00}&=\frac{1}{2}\left(-\partial^2h_{00}+2\partial_v\partial_k h_{0k}- \partial_v^2 h_{kk}\right),\\
R_{0i}&=\frac{1}{2}\left(-\partial^2 h_{0i}+\partial_i\partial_k h_{0k}+ \partial_v \partial_k h_{ik}-\partial_v \partial_i h_{kk}\right),\\
R_{ij}&=\frac{1}{2}\left[-\partial^2 h_{ij}+\left(\partial_i\partial_k h_{jk} +\partial_j \partial_k h_{ik} \right)+\partial_i\partial_j h_{00}\right.\\
&~~~~~~~~\left.-\partial_v\left(\partial_i h_{0j}+\partial_j h_{0i}\right)+\partial_v^2 h_{ij} -\partial_i\partial_j h_{kk}\right],\\
R&=\partial^2h_{00}-\partial^2h_{kk}+\partial_k\partial_l h_{kl}-2\partial_v\partial_k h_{0k} +\partial_v^2h_{kk},
\end{split}
\end{equation}
which were used to compute the logarithmically divergent piece $\tilde{T}_{\mu\nu}^a$. The Weyl tensor is defined as
\begin{equation}
C_{\lambda\mu\sigma\nu}=R_{\lambda\mu\sigma\nu}-\frac{1}{2}\left(R_{\lambda\sigma} g_{\mu\nu} -R_{\lambda\nu}g_{\mu\sigma} -R_{\mu\sigma}g_{\lambda\nu} +R_{\mu\nu} g_{\lambda\sigma}\right)+\frac{1}{6}R \left(g_{\lambda\sigma}g_{\mu\nu}-g_{\lambda\nu} g_{\mu\sigma} \right).
\end{equation}
In terms of the basic structures listed in Table~\ref{table1}, the tensors introduced in~(\ref{diss}) are
\begin{equation}\label{tensor strucutres}
\begin{split}
\nabla_{\langle i}u_{j\rangle}&=\hat{\sigma}_{ij}-\tilde{\sigma}_{ij}+\frac{1}{2} \partial_v \tau_{ij},~~~~
\nabla_{\langle i}\nabla_{j\rangle}\nabla u=\hat{\pi}_{ij}-\tilde{\pi}_{ij}+\frac{1}{2} \partial_v \chi_{ij},\\
u^{\alpha}u^{\beta}C_{\langle i\alpha j\rangle\beta}&=-\frac{1}{4}\left(\partial^2 \tau_{ij} + \partial_v^2 \tau_{ij}-2 \psi_{ij}+\varphi_{ij}+\chi_{ij}+2\partial_v \tilde{\sigma}_{ij}\right),\\
u^{\alpha}\nabla^{\beta}C_{\langle i\alpha j\rangle\beta}&=\frac{1}{2}\left[-2\partial_v C_{i0j0} +\partial_k\left(C_{i0jk}+C_{ikj0}\right)\right]\\
&=\frac{1}{4}\left(\partial_v^3 \tau_{ij} -\partial_v\partial^2\tau_{ij} + \partial_v \psi_{ij} +\partial_v \varphi_{ij}-2\partial_v^2\tilde{\sigma}_{ij} + \partial^2 \tilde{\sigma}_{ij}- \tilde{\pi}_{ij}\right),\\
\nabla^{\alpha}\nabla^{\beta}C_{\langle i \alpha j\rangle\beta}&=\partial_v^2C_{i0j0} - \partial_v \partial_k\left(C_{i0jk}+C_{ikj0}\right) +\partial_k\partial_l C_{\langle ikj\rangle l}\\
&=\frac{1}{12}\left(6\partial_v^2\partial^2 \tau_{ij}-3\partial_v^4\tau_{ij} -3\partial^4 \tau_{ij} -6\partial_v^2\psi_{ij} -3\partial_v^2 \varphi_{ij}+ \partial_v^2 \chi_{ij}+ 6\partial_v^3\tilde{\sigma}_{ij} \right.\\
&~~~~~~~~~\left.-6\partial_v\partial^2\tilde{\sigma}_{ij} +4 \partial_v \tilde{\pi}_{ij} +\partial^2 \varphi_{ij}-\partial^2 \chi_{ij}+ 6 \partial^2 \psi_{ij}-2\phi_{ij}\right),\\
u^{\alpha}\nabla_{\alpha}R_{\langle ij\rangle}&=\frac{1}{2} \left(\partial_v^3\tau_{ij} -\partial_v\partial^2 \tau_{ij}+ 2 \partial_v \psi_{ij} +\partial_v \varphi_{ij}-2 \partial_v^2\tilde{\sigma}_{ij}-\partial_v \chi_{ij}\right),
\end{split}
\end{equation}
%\vspace{0.1em}
where the linearization approximation~(\ref{linearization}) has been applied.

\section{Weyl-covariant formalism for relativistic conformal fluid}\label{appendixw}
In this appendix, we briefly summarize the Weyl-covariant formalism of~\cite{0801.3701}. The main ingredient is the Weyl-covariant derivative operator $\mathcal{D}_{\lambda}$. Consider a Weyl-covariant tensor $Q_{\nu\cdots}^{\mu\cdots}$ which transforms homogenously under the Weyl rescaling,
\begin{equation}
Q_{\nu\cdots}^{\mu\cdots}=e^{-w\phi}\tilde{Q}_{\nu\cdots}^{\mu\cdots}~~~\textrm{under}~~~ g_{\mu\nu}=e^{2\phi}\tilde{g}_{\mu\nu}.
\end{equation}
The Weyl-covariant derivative is defined as
\begin{equation}\label{weyl derivative}
\begin{split}
\mathcal{D}_{\lambda}Q_{\nu\cdots}^{\mu\cdots}\equiv&\nabla_{\lambda}Q_{\nu\cdots}^{\mu\cdots} +w\,\mathcal{A}_{\lambda}Q_{\nu\cdots}^{\mu\cdots}+\left[g_{\lambda\alpha} \mathcal{A}^{\mu}-\delta^{\mu}_{\lambda}\mathcal{A}_{\alpha}-\delta_{\alpha}^{\mu}\right] Q_{\nu\cdots}^{\alpha\cdots}+\cdots\\
&-\left[g_{\lambda\nu}\mathcal{A}^{\alpha}-\delta_{\lambda}^{\alpha}\mathcal{A}_{\nu} -\delta_{\nu}^{\alpha}\mathcal{A}_{\lambda}\right]Q_{\alpha\cdots}^{\mu\cdots}-\cdots,
\end{split}
\end{equation}
where $\mathcal{A}_{\mu}$ is the Weyl-covariant connection and is defined as
\begin{equation}
\mathcal{A}_{\mu}\equiv u^{\alpha}\nabla_{\alpha}u_{\mu}-\frac{1}{3}u_{\mu} \nabla_{\alpha} u^{\alpha}.
\end{equation}
Under the Weyl rescaling, $\mathcal{A}_{\mu}$ transforms in the same way as  $U(1)$ connection,
\begin{equation}
\mathcal{A}_{\mu}=\tilde{\mathcal{A}}_{\mu}+\partial_{\mu}\phi,~~~\textrm{under}~~~ g_{\mu\nu}=e^{2\phi}\tilde{g}_{\mu\nu}.
\end{equation}
Notice that under linearization, $\mathcal{A}\sim \partial_0 u$ and hence any term which is being multiplied by $\mathcal{A}$ is formally non-linear. The Weyl-covariant derivative $\mathcal{D}_{\lambda}$ is metric compatible $\mathcal{D}_{\lambda} g_{\mu\nu}=0$. The most important feature of $\mathcal{D}_{\lambda}$ is that it commutes with the Weyl transformation,
\begin{equation}\label{commu}
\mathcal{D}_{\lambda}Q_{\nu\cdots}^{\mu\cdots}=e^{-w\phi}\tilde{\mathcal{D}}_{\lambda} \tilde{Q}_{\nu\cdots}^{\mu\cdots}~~~\textrm{under}~~~ g_{\mu\nu}=e^{2\phi} \tilde{g}_{\mu\nu}.
\end{equation}

\section{$\tilde{T}_{\mu\nu}(r)$ in terms of metric corrections}\label{appendix2}
In this appendix, we provide the expression of $\tilde{T}_{\mu\nu}(r)$. In terms of the metric corrections, the first piece $\tilde{T}_{\mu\nu}^{n}(r)$ is
\begin{equation} \label{stress tensor1}
\left\{
\begin{aligned}
\tilde{T}_{00}^{n}=&3-12\epsilon {\bf b}_1-3\epsilon h_{00}
+\epsilon\left(\frac{1}{2}r^2 \partial_i\partial_j h_{ij}-\frac{1}{2}r^2\partial^2 h_{kk}-r^2\partial^2h+\frac{1}{2}r^2\partial_i\partial_j \alpha_{ij}\right.\\
&\left.-9r^4h+3k-2r^3\partial u+2r^3\partial_kh_{0k}-r^3\partial_vh_{kk}+\frac{2}{r} \partial j-3r^3\partial_vh-3r^5\partial_rh\right),\\
\tilde{T}_{0i}^{n}=&-4\epsilon u_i+\epsilon h_{0i}
+\epsilon\left(\frac{1}{2}r^3\partial_i h_{00}+\frac{1}{r}\partial_ik+4j_i-r\partial_r j_i-r^3\partial_vu_i-\frac{3}{2}r^3\partial_ih\right.\\
&\left.-\frac{1}{2}r^2\partial^2h_{0i}+\frac{1}{2}r^2\partial_i\partial_kh_{0k} + \frac{1}{2}r^2\partial_v\partial_kh_{ik}-\frac{1}{2}r^2\partial_v\partial_i h_{kk}
- \frac{1}{2r^2}\partial^2 j_i\right.\\
&\left.+\frac{1}{2r^2}\partial_i\partial j+\frac{1}{2}r^2\partial_v\partial_k\alpha_{ik} -r^2\partial_v\partial_ih \right),\\
\tilde{T}_{ij}^{n}=&\delta_{ij}\left(1-4\epsilon {\bf b}_1\right)+\epsilon h_{ij}+\epsilon \delta_{ij} \left[-\frac{1}{2}r^2\partial^2h_{00}+ \frac{1}{2}r^2\partial^2h_{kk}- \frac{1}{2} r^2\partial_k\partial_lh_{kl}\right.\\
&\left.+r^2\partial_v\partial_kh_{0k} -\frac{1}{2}r^2 \partial_v^2 h_{kk}+ \frac{1}{2}r^2\partial^2h-\frac{1}{2r^2} \partial^2k-\frac{r^2}{2} \partial_k\partial_l\alpha_{kl}+\frac{1}{r^2}\partial_v\partial j\right.\\
&\left.-r^2\partial_v^2h+9r^4h+k+2r^3\partial u-2r^3\partial_kh_{0k}+ r^3\partial_v h_{kk}-\frac{2}{r}\partial j -r^3\partial_vh\right.\\
&\left.+ \frac{1}{r}\partial_vk +2r^5\partial_r h-r\partial_r k\right]
+\epsilon\left[\frac{1}{2}r^2\partial_i\partial_jh_{00}-\frac{1}{2}r^2\partial^2h_{ij} -\frac{1}{2}r^2\partial_i\partial_jh_{kk}\right.\\
&\left.+\frac{1}{2}r^2\left(\partial_i\partial_kh_{jk}+ \partial_j\partial_kh_{ik}\right)-\frac{1}{2}r^2\partial_v\left(\partial_ih_{0j}+ \partial_jh_{0i} \right) +\frac{1}{2}r^2\partial_v^2 h_{ij}\right.\\
&\left.-\frac{1}{2}r^2 \partial_i\partial_jh+ \frac{1}{2r^2}\partial_i\partial_jk- \frac{1}{2}r^2 \partial^2\alpha_{ij}
+\frac{1}{2} r^2\left(\partial_i\partial_k\alpha_{jk}+\partial_j\partial_k \alpha_{ik}\right) \right.\\
&\left.-\frac{1}{2r^2}\partial_v\left(\partial_ij_j+\partial_j j_i\right) +\frac{1}{2}r^2 \partial_v^2\alpha_{ij}-r^3\left(\partial_iu_j+ \partial_j u_i\right)+r^3\left(\partial_i h_{0j} + \partial_j h_{0i}\right)  \right.\\
&\left.-r^3\partial_vh_{ij}+ \frac{1}{r} \left(\partial_i j_j +\partial_j j_i\right)-r^3 \partial_v \alpha_{ij} - r^5\partial_r\alpha_{ij}\right],
\end{aligned} \right.
\end{equation}
where it is important to notice that the terms that explicitly vanish at $r=\infty$ have been ignored. We also need the expression for $\tilde{T}_{\mu\nu}^a(r)$
\begin{equation}\label{stress tensor2}
\left\{
\begin{aligned}
\tilde{T}_{00}^a=&\frac{\log{r}}{12}\epsilon\left(-2\partial^2h_{00}+\partial_v^2\partial^2 h_{kk} -\partial^4 h_{kk}-3\partial_v^2\partial_k\partial_l h_{kl}+\partial^2 \partial_k\partial_l h_{kl}\right.\\
&~~~~~~~~~\left.+4\partial_v\partial^2 \partial_kh_{0k} \right),\\
\tilde{T}_{0i}^a=&\frac{\log{r}}{12}\epsilon\left\{3\left(\partial_v^2\partial^2- \partial^4\right) h_{0i}-2\partial_v\partial^2\partial_ih_{00}+\left(\partial_v^2+3 \partial^2 \right) \partial_i\partial_k h_{0k}\right.\\
&~~~~~~~~~\left.+\left(\partial_v^3-\partial_v\partial^2\right)\partial_i h_{kk} - 2\partial_v\partial_i\partial_k\partial_l h_{kl}-3\left(\partial_v^3- \partial_v \partial^2\right)\partial_k h_{ik}\right\},\\
\tilde{T}_{ij}^a=&\frac{\log{r}}{36}\epsilon\left\{3\left(\partial^2-3\partial_v^2\right) \varphi_{ij} +3\left(\partial_v^2-\partial^2\right)\chi_{ij}-6\phi_{ij}+12 \partial_v\tilde{\pi}_{ij}\right.\\
&~~~~~~~~~\left.+18\partial_{v}\left(\partial_v^2-\partial^2\right)\tilde{\sigma}_{ij}+ 18 \left(\partial^2-\partial_v^2\right)\psi_{ij}-9\left(\partial_v^2-\partial^2\right)^2 \tau_{ij}\right.\\
&~~~~~~~~~\left.+\delta_{ij}\left[2\partial^2h_{00}-\left(\partial^2-\partial_v^2\right) \partial^2 h_{kk} +\left(\partial^2-3\partial_v^2\right)\partial_k\partial_l h_{kl} \right] \right\}.
\end{aligned}
\right.
\end{equation}
Definition of the tensors $\varphi_{ij}$ etc. can be found in the Table~\ref{table1}.

\section{Derivation of constraints~(\ref{constraint ti})}\label{appendix3}
In this appendix, we derive the constraints~(\ref{constraint ti}) by suitably combining the differential equations of subsection~\ref{subsection41}. Consider the variables
\begin{align}
X_1&\equiv T_9-i\omega T_3-q^2T_7,~~~~~~~~~~~~~~~~~~~~~~~~~~~Y_1\equiv -i\omega V_3-q^2V_7,\nonumber\\
X_2&\equiv 2\left(T_1+T_3\right)-2i\omega T_5-q^2\left(T_2+T_4\right),~~~~
Y_2\equiv 2\left(V_1+V_3\right)-4i\omega V_5-2q^2\left(V_2+V_4\right),\nonumber\\
X_3&\equiv 2\left(T_6+T_7\right)-i\omega T_4-2q^2T_8,~~~~~~~~~~~~~~Y_3\equiv 2V_6+V_7-i\omega V_4-2q^2V_8,\\
X_4&\equiv T_5-T_6-i\omega T_4-q^2 T_8,~~~~~~~~~~~~~~~~~~~~Y_4\equiv V_5-V_6-i\omega V_4-q^2 V_8 +\frac{1}{2}r^3. \nonumber
\end{align}
These new variables obey similar equations as $T_i$'s and $V_i$'s,
\begin{equation}\label{xy1}
\left\{
\begin{aligned}
0=&\,(r^7-r^3)\partial_r^2X_1+(5r^6-r^2)\partial_rX_1-2i\omega r^5\partial_rX_1-3i\omega r^4 X_1+Y_1-r\partial_rY_1, \\
0=&\,r\partial_r^2Y_1-3\partial_rY_1-q^2r^3\partial_rX_1.
\end{aligned} \right.
\end{equation}
\begin{equation}\label{xy2}
\left\{
\begin{aligned}
0=&\,(r^7-r^3)\partial_r^2X_2+(5r^6-r^2)\partial_rX_2-2i\omega r^5\partial_rX_2-3i\omega r^4 X_2+\frac{1}{3}q^2r^3X_2\\
&+Y_2-r\partial_rY_2, \\
0=&\,r\partial_r^2Y_2-3\partial_rY_2-\frac{4}{3}q^2r^3\partial_rX_2.
\end{aligned} \right.
\end{equation}
\begin{equation}\label{xy3}
\left\{
\begin{aligned}
0=&\,(r^7-r^3)\partial_r^2X_3+(5r^6-r^2)\partial_rX_3-2i\omega r^5\partial_rX_3-3i\omega r^4 X_3+\frac{1}{3}q^2r^3X_3\\
&+2\left(Y_3-r\partial_rY_3\right), \\
0=&\,r\partial_r^2Y_3-3\partial_rY_3-\frac{2}{3}q^2r^3\partial_rX_3.
\end{aligned} \right.
\end{equation}
\begin{equation}\label{xy4}
\left\{
\begin{aligned}
0=&\,(r^7-r^3)\partial_r^2X_4+(5r^6-r^2)\partial_rX_4-2i\omega r^5\partial_rX_4-3i\omega r^4 X_4+\frac{1}{3}q^2r^3X_4\\
&+2\left(Y_4-r\partial_rY_4\right), \\
0=&\,r\partial_r^2Y_4-3\partial_rY_4-\frac{2}{3}q^2r^3\partial_rX_4.
\end{aligned}
\right.
\end{equation}
Notice that the equations of $X_i$'s and $Y_i$'s are \emph{homogeneous}. Under the boundary conditions summarized in section~\ref{section3}, the functions $X_i$'s and $Y_i$'s have only trivial solutions just like the function $h$. Furthermore, the identity $Y_i=0$ does not give non-trivial constraint relation. Therefore, we arrive at
\begin{equation}
X_1=X_2=X_3=X_4=0,
\end{equation}
which result in the constraint relations~(\ref{constraint ti}).
\section{Navier-Stokes equations and correlation functions}\label{appendix4}
In this appendix, we first show that the conservation laws $\nabla^{\mu}T_{\mu\nu}=0$  emerge from the constraint components of~(\ref{einstein eq}). We find it more convenient to study certain mixtures of the constraints with the dynamical equations. Particularly, the combination $r^2f(r)E_{vr}+E_{vv}=0$ is
\begin{equation}\label{constraint1}
\begin{split}
\partial_v{\bf b}_1=&\frac{1}{3}\partial u-\frac{1}{3}\partial_kh_{0k}+\frac{1}{3r} \partial^2 {\bf b}_1+\frac{1}{24r} \partial^2h_{00}-\frac{r^3}{24}\partial^2 h_{00} + \frac{1}{6}\partial_v h_{kk}\\
&-\left(\frac{1}{12r}+\frac{1}{12}r^3\right)\partial_v\partial u+\frac{1}{6}r^3\partial_v \partial_kh_{0k}-\frac{1}{12}r^3\partial_v^2h_{kk}-\frac{1}{3}\partial j\\
&-\frac{1}{12r}\partial^2k+\frac{1}{4}\partial_vk+\frac{1}{6r}\partial_v\partial j-\frac{1}{12}\left(\frac{1}{r^3}-r\right)\partial_r\partial j,
\end{split}
\end{equation}
whose large $r$ limit yields
\begin{equation}\label{ns1}
\partial_v{\bf b}_1=\frac{1}{3}\partial u-\frac{1}{3}\partial_kh_{0k}+\frac{1}{6} \partial_v h_{kk}.
\end{equation}
The combination $r^2f(r)E_{ri}+E_{vi}=0$ leads to
\begin{equation}\label{constraint2}
\begin{split}
\partial_i{\bf b}_1=&\partial_v u_i+\frac{1}{4}r^3\left(\partial_i\partial u-\partial^2 u_i\right)+\frac{1}{4}r^3\left(\partial^2h_{0i}-\partial_i\partial_jh_{0j}\right)- \frac{1}{2}\partial_ih_{00}- \frac{1}{8}r^3\partial_v\partial_ih_{00}\\
&+\frac{1}{4}r^3\left(\partial_v\partial_ih_{kk}-\partial_v\partial_jh_{ij}\right)+ \frac{1}{4}r^3\partial_v^2u_i+\frac{1}{4r} \left(\partial^2 j_i-\partial_i\partial j\right)+\frac{1}{4} \partial_ik-\partial_vj_i\\
&-\frac{1}{4}r^3 \partial_v\partial_j\alpha_{ij}+ \frac{1}{4}(r-r^5)\partial_r\partial_j \alpha_{ij} -\frac{1}{4}r\partial_r\partial_ik+ \frac{1}{4}r\partial_r\partial_vj_i.
\end{split}
\end{equation}
Taking the large $r$ limit of~(\ref{constraint2}), we arrive at
\begin{equation}\label{ns2}
\begin{split}
\partial_i{\bf b}_1=&\partial_v u_i-\frac{1}{2}\partial_ih_{00} +\frac{1}{3}t_1 (\partial_i\partial u+3\partial^2 u_i) +\frac{2}{3}t_2\partial^2\partial_i\partial u\\
&+\frac{1}{192}\left(192 t_3+3\partial_v\partial^2-7\partial_v^3\right)\partial^2h_{0i} \\
&+\frac{1}{1728}\left(576t_3+1152t_4\partial^2-21\partial_v^3-31\partial_v\partial^2\right) \partial_i\partial_kh_{0k}\\
&+\frac{1}{864}\left(576t_5+21\partial_v^2+\partial^2\right)\partial^2\partial_ih_{00}\\
&+\frac{1}{192}\left(192t_9+192t_7\partial^2+7\partial_v^4-3\partial_v^2\partial^2\right) \partial_kh_{ik}\\
&+\frac{1}{1728}\left(576t_7+1152t_8\partial^2-\partial^2+9\partial_v^2\right) \partial_i \partial_k\partial_lh_{kl}\\
&+\frac{1}{1728}\left[1152t_6\partial^2-576t_9+\left(\partial^2-\partial_v^2\right) \left(21\partial_v^2+\partial^2 \right)\right]\partial_ih_{kk}.
\end{split}
\end{equation}
It can be checked that~(\ref{ns1},\ref{ns2}) are equivalent to the conservation laws $\nabla^{\mu}T_{\mu\nu}=0$ with $T_{\mu\nu}$ given by~(\ref{emt1}). Equations~(\ref{ns1},\ref{ns2}) are usually referred to as the Navier-Stokes equations for hydrodynamics, which determine the time evolution of the temperature and velocity fields once appropriate initial data is specified.

From the hydro-like constitutive relation~(\ref{emt1}), it is straightforward to construct two-point retarded Green's functions for $T_{\mu\nu}$, which characterize the linear response of the fluid system slightly disturbed  from its equilibrium state. In parallel with~\cite{0905.4069}, we present them by different channels, as classified according to the $SO(3)$ symmetry.
\paragraph{Scalar channel}
To study the correlation function in the scalar channel, we just need to turn on the perturbation $h_{xy}(v,z)$ and particularly the fluid is at rest $u_{i}=0$. Then, the component $T^{xy}$ is read off from~(\ref{emt1}),
\begin{equation}
\begin{split}
T^{xy}=&-Ph_{xy}+P\left\{i\omega \eta+\frac{1}{4}\left(\omega^2+q^2\right) \kappa +\frac{1}{4}i\omega (\omega^2-q^2)\rho\right.\\
&~~~~~~~~~~~~~~~~~~\left.- \frac{1}{2}i\omega \left(\omega^2-q^2 \right) \theta-\frac{1}{4} (\omega^2-q^2)^2\xi \right\}h_{xy}.
\end{split}
\end{equation}
So, the retarded Green's function in the scalar channel is
\begin{equation}
\begin{split}\label{scalar channel}
G^{xy,xy}_{R}=&-P+P\left\{i\omega \eta+\frac{1}{4}\left(\omega^2+q^2\right) \kappa +\frac{1}{4}i\omega (\omega^2-q^2)\rho\right.\\
&~~~~~~~~~~~~~~\left.- \frac{1}{2} i\omega \left(\omega^2-q^2\right)\theta-\frac{1}{4} (\omega^2-q^2)^2\xi \right\}.
\end{split}
\end{equation}
\paragraph{Shear channel}
It is sufficient to consider the metric perturbation $h_{0x}(v,z)$, which stirs the fluid to be of velocity $u_{x}(v,z)$. From the Navier-Stokes equations~(\ref{ns1},\ref{ns2}),
\begin{equation}
u_x=-\left[\frac{i\omega q^2(7\omega^2-3q^2)}{192}-q^2t_3\right]\frac{h_{0x}} {-i\omega-q^2t_1}.
\end{equation}
Therefore, the component $T^{0x}$ is
\begin{equation}
T^{0x}=\left(\varepsilon+P\right)u^0u^x+Pg^{0x}=\left(\varepsilon+P\right)(-h_{0x}+u_x)+ Ph_{0x}
\end{equation}
which results in the retarded Green's function in the shear channel
\begin{equation}\label{shear channel}
G_{R}^{0x,0x}=\left(\varepsilon+P\right)\frac{8q^2\eta-2i\omega q^2 \kappa- q^2 \left(q^2- 2\omega^2 \right) \rho-4 \omega^2q^2 \theta -2i\omega q^2 \left(q^2-\omega^2\right) \xi}{32\left(-i\omega+q^2\eta/4\right)}-\varepsilon
\end{equation}
where the pole $-i\omega+q^2\eta/4=0$ is  the shear dispersion relation.
\paragraph{Sound channel}
The metric perturbation which generates the sound mode is $h_{0z}(v,z)$. The conservation laws~(\ref{ns1},\ref{ns2}) force a relation between the fluid velocity $u_{z}(v,z)$ and the metric perturbation
\begin{equation}
\begin{split}
u_z=&\left\{q^2+ i\omega\left[\frac{1}{64}i\omega q^2\left(7\omega^2-3q^2\right)-3q^2 t_3\right] \right.\\
&\left.-i\omega q^2\left[-\frac{1}{16\times 36}i\omega(21\omega^2+31q^2)+ \left(t_3-2q^2t_4\right) \right]\right\}\\
&\times \frac{1}{q^2-3\omega^2-i\omega q^2 \eta+i\omega q^4\zeta/2}h_{0z}.
\end{split}
\end{equation}
Thus, the component $T^{0z}$ is computed as
\begin{equation}
T^{0z}=\left(\varepsilon+P\right)u^0u^z+Pg^{0z}=\left(\varepsilon+P\right)\left(-h_{0z}+ u_z\right)+Ph_{0z},
\end{equation}
from which we read off the retarded Green's function in the sound channel
\begin{equation}\label{sound channel}
G^{0z,0z}_{R}=(\varepsilon+P)\frac{N_{*}}{q^2-3\omega^2-i\omega q^2\eta+i\omega q^4 \zeta/2} -\varepsilon.
\end{equation}
The numerator $N_{*}$ is
\begin{equation}\nonumber
N_{*}=q^2-i\omega q^2\eta+i\omega q^4 \zeta-\frac{1}{4}\omega^2q^2\kappa
- \frac{1}{4}i\omega ^3 q^2\rho+\frac{1}{2}i\omega^3q^2\theta -\frac{1}{12}\omega^2 q^2\left(q^2-3\omega^2\right) \xi .
\end{equation}
From~(\ref{sound channel}), we recover the sound wave dispersion relation $q^2-3\omega^2-i\omega q^2\eta+i\omega q^4 \zeta/2=0$. Up to normalization convention and the $\theta$-terms, our expressions for two-point correlation functions are the same as those of~\cite{0905.4069}. We also reproduce the plots~\cite{0905.4069} (also~\cite{hep-th/0602059}) for the thermal correlators using our numerical results for the viscosities and GSFs.

Choosing the spatial momentum along $z$-direction, i.e., $\bar{k}_{\mu}=\left(-\bar{\omega},0,0,\bar{q}\right)$, the components of the correlators are conveniently written as
\begin{equation}\label{zero}
\begin{split}
G_R^{xy,xy}&=\frac{1}{2} G_3\left(\bar{\omega},\bar{q}\right),\\
G_R^{0x,0x}&=\frac{1}{2}\frac{\bar{q}^2}{\bar{\omega}^2-\bar{q}^2} G_1\left(\bar{\omega},\bar{q}\right),\\
G_R^{00,00}&=\frac{2}{3}\frac{\bar{q}^4}{\left(\bar{\omega}^2-\bar{q}^2\right)^2} G_2\left(\bar{\omega},\bar{q}\right),\\
G_R^{0z,0z}&=\frac{2}{3}\frac{\bar{\omega}^2\bar{q}^2}{\left(\bar{\omega}^2-\bar{q}^2\right)^2} G_2\left(\bar{\omega},\bar{q}\right),
\end{split}
\end{equation}
where the bar is used to emphasize that the momenta here are dimensional. At zero temperature, the correlators in three symmetry channels coincide $G_1=G_2=G_3=G_s$. For $\mathcal{N}=4$ super-Yang-Mills theory, their expressions can be found in~\cite{hep-th/0205051,hep-th/0602059}
\begin{equation}\label{zero temp}
G_s=\frac{N_c^2\bar{k}^4}{32\pi^2}\left(\ln{|\bar{k}^2|}-i\pi \Theta\left(-\bar{k}^2\right) \textrm{sign}\bar{\omega}\right),
\end{equation}
where the color $N_c$ is related to $G_{N}$ by $1/G_N=2N_c^2/\pi$. We use our numerical results for viscosities and GSFs to construct the thermal correlators and find that they start to approach their zero temperature results~(\ref{zero temp}) near $\omega\sim q\sim 5$.

\section{Perturbative solutions for $V_i$'s and $T_i$'s}\label{appendix5}
In this appendix, we present perturbative solutions for $V_i$'s and $T_i$'s. Recall the formal expansion for the functions $V_i$'s and $T_i$'s in~(\ref{VT exp}),
\begin{equation}
V_i(\omega,q_i,r)=\sum_{n=0}^{\infty}\lambda^nV_i^{(n)}(\omega,q_i,r),~~~
T_i(\omega,q_i,r)=\sum_{n=0}^{\infty}\lambda^nT_i^{(n)}(\omega,q_i,r).
\end{equation}
In what follows, we collect the results according to different sectors.
\paragraph{I:~$\left\{V_1,~T_1,~V_2,~T_2\right\}$}
\begin{equation}
\begin{split}
V_1^{(0)}&=0,~~~V_1^{(1)}=-i\omega r^3,\\
T_1^{(0)}&=\frac{1}{4}\left[\ln{\frac{(1+r^2)(1+r)^2}{r^4}}-2\arctan(r)+\pi\right]
\xlongrightarrow{r\to\infty}\frac{1}{r}-\frac{1}{4r^4}+\mathcal{O}\left(\frac{1}{r^5} \right),\\
V_2^{(0)}&=-\int_{1}^{r}x^3dx\int_{x}^{\infty} \left[\frac{1}{y^3}-\frac{\partial_y T_1^{(0)}(y)}{3y}\right] dy- \frac{3}{8} \xlongrightarrow{r\to\infty} -\frac{1}{3}r^2 +\mathcal{O}\left(\frac{1}{r} \right),\\
T_1^{(1)}&=-i\omega\int_{r}^{\infty}\frac{i\omega dx}{x^5-x}\int_{1}^{x} dy\left[2y^3 \partial_y T_1^{(0)}(y)-y+ 3y^2 T_1^{(0)}(y) \right]\\
&\xlongrightarrow{r\to\infty}-\frac{i\omega}{4r^4}\left(1- \frac{\ln{2}}{2}\right) +\mathcal{O}\left(\frac{1}{r^5}\right),\\
V_1^{(2)}&=\int_{r}^{\infty}x^3dx\int_{x}^{\infty}dy \left[\frac{q^2}{y}\partial_y T_1^{(0)}(y)+\frac{q^2}{y^3}\right]\xlongrightarrow{r\to\infty} \mathcal{O} \left(\frac{1}{r}\right),\\
T_2^{(0)}&=\int_{r}^{\infty}\frac{dx}{x^5-x}\int_{1}^{x}\left[-\frac{2V_2^{(0)}(y)}{y^2} +\frac{2\partial_y V_2^{(0)}(y)}{y}+\frac{2}{3}y T_1^{(0)}(y)\right]dy\\
& \xlongrightarrow{r\to\infty}-\frac{1}{48r^4}\left(5-\pi-2\ln{2}\right) + \mathcal{O}\left(\frac{1}{r^5}\right),\\
V_2^{(1)}&=-\int_{r}^{\infty}x^3dx\int_{x}^{\infty}dy\frac{1}{3y}\partial_yT_1^{(0)}(y) \xlongrightarrow{r\to\infty} \mathcal{O}\left(\frac{1}{r}\right),\\
T_1^{(2)}&=\int_{r}^{\infty}\frac{dx}{x-x^5}\int_{1}^{x}dy\left[2i\omega y^3 \partial_y T_1^{(1)}(y)+ 3i\omega y^2 T_1^{(1)}(y) +\frac{\partial_y V_1^{(2)}(y)}{y}- \frac{V_1^{(2)}(y)}{y^2}\right]\\
&\xlongrightarrow{r\to\infty} \frac{1}{192r^4}\left\{6q^2+\omega^2 \left[6\pi- \pi^2 +12\left(2-3\ln{2}\ln^2{2}\right)\right]\right\}+ \mathcal{O}\left(\frac{1}{r^5}\right).
\end{split}
\end{equation}
\paragraph{II:~$\left\{V_3,~T_3,~V_4,~T_4\right\}$}
\begin{equation}
\begin{split}
V_3^{(0)}&=V_3^{(1)}=0,\\
T_3^{(0)}&=-\frac{1}{4}\left[\ln{\frac{(1+r^2)(1+r)^2}{r^4}}-2\arctan(r)+\pi\right]
\xlongrightarrow{r\to\infty}-\frac{1}{r}+\frac{1}{4r^4}+\mathcal{O}\left(\frac{1}{r^5} \right),\\
V_4^{(0)}&=\int_{1}^{r}x^3dx\int_{x}^{\infty}\frac{\partial_yT_3^{(0)}(y)}{3y} dy+\frac{1}{8}\xlongrightarrow{r\to\infty}\frac{1}{12}r^2+\mathcal{O}\left(\frac{1}{r} \right),\\
T_4^{(0)}&=-\int_{r}^{\infty}\frac{dx}{x^5-x}\int_{1}^{x}\left[-\frac{2V_4^{(0)}(y)}{y^2} +\frac{2\partial_y V_4^{(0)}(y)}{y}+\frac{2}{3}y T_3^{(0)}(y)\right]dy\\
& \xlongrightarrow{r\to\infty}\frac{1}{6r^3}-\frac{1}{48r^4}\left(1+\pi+2\ln{2}\right) + \mathcal{O}\left(\frac{1}{r^5}\right),\\
T_3^{(1)}&=-\int_{r}^{\infty}\frac{i\omega dx}{x^5-x}\int_{1}^{x} dy\left[2y^3 \partial_y T_3^{(0)}(y)+ 3y^2 T_3^{(0)}(y) \right]\\
&\xlongrightarrow{r\to\infty}\frac{i\omega}{4r^2}-\frac{i\omega}{8r^4}\left(-1+ \ln{2}\right)+\mathcal{O}\left(\frac{1}{r^5}\right).
\end{split}
\end{equation}
\paragraph{III:~$\left\{V_5,~T_5\right\}$}
\begin{equation}
\begin{split}
V_5^{(0)}&=-\frac{1}{2}r^3,~~~V_5^{(1)}=0,\\
T_5^{(0)}&=\frac{1}{4}\ln{\frac{1+r^2}{r^2}}\xlongrightarrow{r\to\infty} \frac{1}{4r^2} -\frac{1}{8r^4}+\mathcal{O}\left(\frac{1}{r^5}\right),\\
T_5^{(1)}&=-\int_{r}^{\infty}\frac{i\omega dx}{x^5-x}\int_1^x dy\left[ 2y^3 \partial_y T_5^{(0)}(y) + 3y^2 T_5^{(0)}(y)\right]\\
&\xlongrightarrow{r\to \infty}\frac{i\omega}{12r^3}-\frac{i\omega}{32r^4}\left(4+\pi -2\ln{2}\right)+\mathcal{O}\left(\frac{1}{r^5}\right).
\end{split}
\end{equation}
\paragraph{IV:~$\left\{T_9\right\}$}
\begin{equation}
\begin{split}
T_9^{(0)}&=0,\\
T_9^{(1)}&=-\frac{1}{4}i\omega \left[\ln{\frac{(1+r^2)(1+r)^2}{r^4}}-2\arctan(r) +\pi\right]\\
&\xlongrightarrow{r\to\infty}-\frac{i\omega}{r}+\frac{i\omega}{4r^4}+\mathcal{O} \left(\frac{1}{r^5} \right),\\
T_9^{(2)}&=-\int_r^{\infty}\frac{dx}{x^5-x}\int_1^xdy \left[2i\omega y^3 \partial_y T_9^{(1)}(y) + 3i\omega y^2 T_9^{(1)}(y)+q^2 y\right]\\
&\xlongrightarrow{r\to\infty} -\frac{q^2+\omega^2}{4r^2}+\frac{1}{8r^4}\left[q^2+ \left(\ln{2}-1\right)\omega^2\right]+\mathcal{O}\left(\frac{1}{r^5}\right).
\end{split}
\end{equation}
\paragraph{V,VI,VII:~$\left\{V_6,~T_6,~V_7,~T_7,~V_8,~T_8\right\}$}
\begin{equation}
\begin{split}
V_6^{(0)}&=0,~~~V_7^{(0)}=0,\\
T_6^{(0)}&=-T_7^{(0)}=\frac{1}{4}\ln{\frac{1+r^2}{r^2}}\xlongrightarrow{r\to\infty} \frac{1}{4r^2}-\frac{1}{8r^4}+ \mathcal{O}\left(\frac{1}{r^5}\right),\\
V_8^{(0)}&=\int_1^rx^3dx\int_r^{\infty}dy\frac{\partial_y T_7^{(0)}(y)}{3y}+ \frac{1}{12} \xlongrightarrow{r\to\infty} \frac{1}{18}r+\mathcal{O}\left(\frac{1}{r}\right),\\
T_8^{(0)}&=-\int_r^{\infty}\frac{dx}{x^5-x}\int_1^x dy\left[-\frac{2V_8^{(0)}(y)}{y^2} +\frac{2\partial_y V_8^{(0)}(y)}{y}+\frac{2}{3}yT_7^{(0)}(y)\right]\\
&\xlongrightarrow{r\to\infty}\frac{13-12\ln{2}}{288r^4}+\frac{\log{r}}{24r^4}+\mathcal{O} \left(\frac{\log{r}}{r^5}\right).
\end{split}
\end{equation}
The power expansion of $t_i$'s can be directly read off from these integrals. In principle, one can extend the above perturbative analysis to arbitrary order in derivative expansion. Then one can write down recursive relations for transport coefficients among different orders in the derivative expansion, as done in~\cite{1409.3095}.

\acknowledgments
ML thanks Edward Shuryak for  early collaborative work that lead to this project. ML is also grateful to the Physics Department of the University of Connecticut for hospitality during the period when this publication was completed.
This work was supported by the ISRAELI SCIENCE FOUNDATION grant \#87277111, BSF grant \#012124, the People Program (Marie Curie Actions) of the European Union's Seventh Framework under REA grant agreement \#318921; and the Council for Higher Education of Israel under the PBC Program of Fellowships for Outstanding Post-doctoral Researchers from China and India (2013-2014).

\providecommand{\href}[2]{#2}\begingroup\raggedright\endgroup

%\bibliographystyle{utphys}
%\bibliography{reference}

\begin{thebibliography}{10}

\bibitem{1406.7222}
Y.~Bu and M.~Lublinsky, ``{All Order Linearized Hydrodynamics from
  Fluid/Gravity Correspondence},''
  \href{http://dx.doi.org/10.1103/PhysRevD.90.086003}{{\em Phys.Rev.}
  {\bfseries D90} (2014) 086003},
\href{http://arxiv.org/abs/1406.7222}{{\ttfamily arXiv:1406.7222 [hep-th]}}.
%%CITATION = ARXIV:1406.7222;%%.

\bibitem{1409.3095}
Y.~Bu and M.~Lublinsky, ``{Linearized fluid/gravity correspondence: from shear
  viscosity to all order hydrodynamics},''
  \href{http://dx.doi.org/10.1007/JHEP11(2014)064}{{\em JHEP} {\bfseries 1411}
  (2014) 064},
\href{http://arxiv.org/abs/1409.3095}{{\ttfamily arXiv:1409.3095 [hep-th]}}.
%%CITATION = ARXIV:1409.3095;%%.

\bibitem{0905.4069}
M.~Lublinsky and E.~Shuryak, ``{Improved Hydrodynamics from the AdS/CFT},''
  \href{http://dx.doi.org/10.1103/PhysRevD.80.065026}{{\em Phys.Rev.}
  {\bfseries D80} (2009) 065026},
\href{http://arxiv.org/abs/0905.4069}{{\ttfamily arXiv:0905.4069 [hep-ph]}}.
%%CITATION = ARXIV:0905.4069;%%.

\bibitem{fluid1}
L.~D. Landau and E.~M. Lifshitz, {\em Fluid Mechanics: Course of Theoretical
  Physics, Vol. 6}.
\newblock Butterworth-Heinemann, 1965.

\bibitem{km}
L.~P. Kadanoff and P.~C. Martin, ``Hydrodynamic equations and correlation
  functions,'' \href{http://dx.doi.org/10.1016/0003-4916(63)90078-2}{{\em
  Annals Phys.} {\bfseries 24} (1963) 419--469}.

\bibitem{fluid2}
D.~Forster, {\em Hydrodynamic Fluctuations, Broken Symmetry, and Correlation
  Functions}.
\newblock Westview Press, 1995.

\bibitem{Muller}
I.~Muller, ``{Zum Paradoxon der Warmeleitungstheorie},''
\href{http://dx.doi.org/10.1007/BF01326412}{{\em Z.Phys.} {\bfseries 198}
  (1967) 329--344}.
%%CITATION = ZEPYA,198,329;%%.

\bibitem{Israel}
W.~Israel, ``{Nonstationary irreversible thermodynamics: A Causal relativistic
  theory},''
\href{http://dx.doi.org/10.1016/0003-4916(76)90064-6}{{\em Annals Phys.}
  {\bfseries 100} (1976) 310--331}.
%%CITATION = APNYA,100,310;%%.

\bibitem{IS1976}
W.~Israel and J.~Stewart, ``{Thermodynamics of nonstationary and transient
  effects in a relativistic gas},''
\href{http://dx.doi.org/10.1016/0375-9601(76)90075-X}{{\em Phys. Lett.}
  {\bfseries A 58} (1976) 213--215}.
%%CITATION = APNYA,118,341;%%.

\bibitem{IS1979}
W.~Israel and J.~Stewart, ``{Transient relativistic thermodynamics and kinetic
  theory},''
\href{http://dx.doi.org/10.1016/0003-4916(79)90130-1}{{\em Annals Phys.}
  {\bfseries 118} (1979) 341--372}.
%%CITATION = APNYA,118,341;%%.

\bibitem{Hiscock1983}
W.~Hiscock and L.~Lindblom, ``{Stability and causality in dissipative
  relativistic fluids},''
\href{http://dx.doi.org/10.1016/0003-4916(83)90288-9}{{\em Annals Phys.}
  {\bfseries 151} (1983) 466--496}.
%%CITATION = APNYA,151,466;%%.

\bibitem{Hiscock1985}
W.~A. Hiscock and L.~Lindblom, ``{Generic instabilities in first-order
  dissipative relativistic fluid theories},''
\href{http://dx.doi.org/10.1103/PhysRevD.31.725}{{\em Phys.Rev.} {\bfseries
  D31} (1985) 725--733}.
%%CITATION = PHRVA,D31,725;%%.

\bibitem{Hiscock1987}
W.~A. Hiscock and L.~Lindblom, ``{Linear plane waves in dissipative
  relativistic fluids},''
\href{http://dx.doi.org/10.1103/PhysRevD.35.3723}{{\em Phys.Rev.} {\bfseries
  D35} (1987) 3723--3732}.
%%CITATION = PHRVA,D35,3723;%%.

\bibitem{0807.3120}
G.~Denicol, T.~Kodama, T.~Koide, and P.~Mota, ``{Stability and Causality in
  relativistic dissipative hydrodynamics},''
  \href{http://dx.doi.org/10.1088/0954-3899/35/11/115102}{{\em J.Phys.}
  {\bfseries G35} (2008) 115102},
\href{http://arxiv.org/abs/0807.3120}{{\ttfamily arXiv:0807.3120 [hep-ph]}}.
%%CITATION = ARXIV:0807.3120;%%.

\bibitem{0907.3906}
S.~Pu, T.~Koide, and D.~H. Rischke, ``{Does stability of relativistic
  dissipative fluid dynamics imply causality?},''
  \href{http://dx.doi.org/10.1103/PhysRevD.81.114039}{{\em Phys.Rev.}
  {\bfseries D81} (2010) 114039},
\href{http://arxiv.org/abs/0907.3906}{{\ttfamily arXiv:0907.3906 [hep-ph]}}.
%%CITATION = ARXIV:0907.3906;%%.

\bibitem{1102.4780}
G.~S. Denicol, J.~Noronha, H.~Niemi, and D.~H. Rischke, ``{Origin of the
  Relaxation Time in Dissipative Fluid Dynamics},''
  \href{http://dx.doi.org/10.1103/PhysRevD.83.074019}{{\em Phys.Rev.}
  {\bfseries D83} (2011) 074019},
\href{http://arxiv.org/abs/1102.4780}{{\ttfamily arXiv:1102.4780 [hep-th]}}.
%%CITATION = ARXIV:1102.4780;%%.

\bibitem{hep-th/9711200}
J.~M. Maldacena, ``{The Large N limit of superconformal field theories and
  supergravity},'' \href{http://dx.doi.org/10.1023/A:1026654312961}{{\em
  Int.J.Theor.Phys.} {\bfseries 38} (1999) 1113--1133},
\href{http://arxiv.org/abs/hep-th/9711200}{{\ttfamily arXiv:hep-th/9711200
  [hep-th]}}.
%%CITATION = HEP-TH/9711200;%%.

\bibitem{hep-th/9802109}
S.~Gubser, I.~R. Klebanov, and A.~M. Polyakov, ``{Gauge theory correlators from
  noncritical string theory},''
  \href{http://dx.doi.org/10.1016/S0370-2693(98)00377-3}{{\em Phys.Lett.}
  {\bfseries B428} (1998) 105--114},
\href{http://arxiv.org/abs/hep-th/9802109}{{\ttfamily arXiv:hep-th/9802109
  [hep-th]}}.
%%CITATION = HEP-TH/9802109;%%.

\bibitem{hep-th/9802150}
E.~Witten, ``{Anti-de Sitter space and holography},'' {\em
  Adv.Theor.Math.Phys.} {\bfseries 2} (1998) 253--291,
\href{http://arxiv.org/abs/hep-th/9802150}{{\ttfamily arXiv:hep-th/9802150
  [hep-th]}}.
%%CITATION = HEP-TH/9802150;%%.

\bibitem{hep-th/0205052}
G.~Policastro, D.~T. Son, and A.~O. Starinets, ``{From AdS / CFT correspondence
  to hydrodynamics},''
  \href{http://dx.doi.org/10.1088/1126-6708/2002/09/043}{{\em JHEP} {\bfseries
  0209} (2002) 043},
\href{http://arxiv.org/abs/hep-th/0205052}{{\ttfamily arXiv:hep-th/0205052
  [hep-th]}}.
%%CITATION = HEP-TH/0205052;%%.

\bibitem{hep-th/0210220}
G.~Policastro, D.~T. Son, and A.~O. Starinets, ``{From AdS / CFT correspondence
  to hydrodynamics. 2. Sound waves},''
  \href{http://dx.doi.org/10.1088/1126-6708/2002/12/054}{{\em JHEP} {\bfseries
  0212} (2002) 054},
\href{http://arxiv.org/abs/hep-th/0210220}{{\ttfamily arXiv:hep-th/0210220
  [hep-th]}}.
%%CITATION = HEP-TH/0210220;%%.

\bibitem{hep-th/0104066}
G.~Policastro, D.~T. Son, and A.~O. Starinets, ``{The Shear viscosity of
  strongly coupled N=4 supersymmetric Yang-Mills plasma},''
  \href{http://dx.doi.org/10.1103/PhysRevLett.87.081601}{{\em Phys.Rev.Lett.}
  {\bfseries 87} (2001) 081601},
\href{http://arxiv.org/abs/hep-th/0104066}{{\ttfamily arXiv:hep-th/0104066
  [hep-th]}}.
%%CITATION = HEP-TH/0104066;%%.

\bibitem{hep-th/0405231}
P.~Kovtun, D.~T. Son, and A.~O. Starinets, ``{Viscosity in strongly interacting
  quantum field theories from black hole physics},''
  \href{http://dx.doi.org/10.1103/PhysRevLett.94.111601}{{\em Phys.Rev.Lett.}
  {\bfseries 94} (2005) 111601},
\href{http://arxiv.org/abs/hep-th/0405231}{{\ttfamily arXiv:hep-th/0405231
  [hep-th]}}.
%%CITATION = HEP-TH/0405231;%%.

\bibitem{hep-th/0311175}
A.~Buchel and J.~T. Liu, ``{Universality of the shear viscosity in
  supergravity},'' \href{http://dx.doi.org/10.1103/PhysRevLett.93.090602}{{\em
  Phys.Rev.Lett.} {\bfseries 93} (2004) 090602},
\href{http://arxiv.org/abs/hep-th/0311175}{{\ttfamily arXiv:hep-th/0311175
  [hep-th]}}.
%%CITATION = HEP-TH/0311175;%%.

\bibitem{0808.1837}
A.~Buchel, R.~C. Myers, M.~F. Paulos, and A.~Sinha, ``{Universal holographic
  hydrodynamics at finite coupling},''
  \href{http://dx.doi.org/10.1016/j.physletb.2008.10.003}{{\em Phys.Lett.}
  {\bfseries B669} (2008) 364--370},
\href{http://arxiv.org/abs/0808.1837}{{\ttfamily arXiv:0808.1837 [hep-th]}}.
%%CITATION = ARXIV:0808.1837;%%.

\bibitem{0809.3808}
N.~Iqbal and H.~Liu, ``{Universality of the hydrodynamic limit in AdS/CFT and
  the membrane paradigm},''
  \href{http://dx.doi.org/10.1103/PhysRevD.79.025023}{{\em Phys.Rev.}
  {\bfseries D79} (2009) 025023},
\href{http://arxiv.org/abs/0809.3808}{{\ttfamily arXiv:0809.3808 [hep-th]}}.
%%CITATION = ARXIV:0809.3808;%%.

\bibitem{0712.2456}
S.~Bhattacharyya, V.~E. Hubeny, S.~Minwalla, and M.~Rangamani, ``{Nonlinear
  Fluid Dynamics from Gravity},''
  \href{http://dx.doi.org/10.1088/1126-6708/2008/02/045}{{\em JHEP} {\bfseries
  0802} (2008) 045},
\href{http://arxiv.org/abs/0712.2456}{{\ttfamily arXiv:0712.2456 [hep-th]}}.
%%CITATION = ARXIV:0712.2456;%%.

\bibitem{0806.4602}
M.~Haack and A.~Yarom, ``{Nonlinear viscous hydrodynamics in various dimensions
  using AdS/CFT},'' \href{http://dx.doi.org/10.1088/1126-6708/2008/10/063}{{\em
  JHEP} {\bfseries 0810} (2008) 063},
\href{http://arxiv.org/abs/0806.4602}{{\ttfamily arXiv:0806.4602 [hep-th]}}.
%%CITATION = ARXIV:0806.4602;%%.

\bibitem{0809.4272}
S.~Bhattacharyya, R.~Loganayagam, I.~Mandal, S.~Minwalla, and A.~Sharma,
  ``{Conformal Nonlinear Fluid Dynamics from Gravity in Arbitrary
  Dimensions},'' \href{http://dx.doi.org/10.1088/1126-6708/2008/12/116}{{\em
  JHEP} {\bfseries 0812} (2008) 116},
\href{http://arxiv.org/abs/0809.4272}{{\ttfamily arXiv:0809.4272 [hep-th]}}.
%%CITATION = ARXIV:0809.4272;%%.

\bibitem{0806.0006}
S.~Bhattacharyya, R.~Loganayagam, S.~Minwalla, S.~Nampuri, S.~P. Trivedi, {\em
  et~al.}, ``{Forced Fluid Dynamics from Gravity},''
  \href{http://dx.doi.org/10.1088/1126-6708/2009/02/018}{{\em JHEP} {\bfseries
  0902} (2009) 018},
\href{http://arxiv.org/abs/0806.0006}{{\ttfamily arXiv:0806.0006 [hep-th]}}.
%%CITATION = ARXIV:0806.0006;%%.

\bibitem{0704.0240}
D.~T. Son and A.~O. Starinets, ``{Viscosity, Black Holes, and Quantum Field
  Theory},''
  \href{http://dx.doi.org/10.1146/annurev.nucl.57.090506.123120}{{\em
  Ann.Rev.Nucl.Part.Sci.} {\bfseries 57} (2007) 95--118},
\href{http://arxiv.org/abs/0704.0240}{{\ttfamily arXiv:0704.0240 [hep-th]}}.
%%CITATION = ARXIV:0704.0240;%%.

\bibitem{0905.4352}
M.~Rangamani, ``{Gravity and Hydrodynamics: Lectures on the fluid-gravity
  correspondence},''
  \href{http://dx.doi.org/10.1088/0264-9381/26/22/224003}{{\em
  Class.Quant.Grav.} {\bfseries 26} (2009) 224003},
\href{http://arxiv.org/abs/0905.4352}{{\ttfamily arXiv:0905.4352 [hep-th]}}.
%%CITATION = ARXIV:0905.4352;%%.

\bibitem{1107.5780}
V.~E. Hubeny, S.~Minwalla, and M.~Rangamani, ``{The fluid/gravity
  correspondence},''
\href{http://arxiv.org/abs/1107.5780}{{\ttfamily arXiv:1107.5780 [hep-th]}}.
%%CITATION = ARXIV:1107.5780;%%.

\bibitem{hep-th/0703243}
M.~P. Heller and R.~A. Janik, ``{Viscous hydrodynamics relaxation time from
  AdS/CFT},'' \href{http://dx.doi.org/10.1103/PhysRevD.76.025027}{{\em
  Phys.Rev.} {\bfseries D76} (2007) 025027},
\href{http://arxiv.org/abs/hep-th/0703243}{{\ttfamily arXiv:hep-th/0703243
  [hep-th]}}.
%%CITATION = HEP-TH/0703243;%%.

\bibitem{0906.4423}
G.~Beuf, M.~P. Heller, R.~A. Janik, and R.~Peschanski, ``{Boost-invariant early
  time dynamics from AdS/CFT},''
  \href{http://dx.doi.org/10.1088/1126-6708/2009/10/043}{{\em JHEP} {\bfseries
  0910} (2009) 043},
\href{http://arxiv.org/abs/0906.4423}{{\ttfamily arXiv:0906.4423 [hep-th]}}.
%%CITATION = ARXIV:0906.4423;%%.

\bibitem{1103.3452}
M.~P. Heller, R.~A. Janik, and P.~Witaszczyk, ``{The characteristics of
  thermalization of boost-invariant plasma from holography},''
  \href{http://dx.doi.org/10.1103/PhysRevLett.108.201602}{{\em Phys.Rev.Lett.}
  {\bfseries 108} (2012) 201602},
\href{http://arxiv.org/abs/1103.3452}{{\ttfamily arXiv:1103.3452 [hep-th]}}.
%%CITATION = ARXIV:1103.3452;%%.

\bibitem{1203.0755}
M.~P. Heller, R.~A. Janik, and P.~Witaszczyk, ``{A numerical relativity
  approach to the initial value problem in asymptotically Anti-de Sitter
  spacetime for plasma thermalization - an ADM formulation},''
  \href{http://dx.doi.org/10.1103/PhysRevD.85.126002}{{\em Phys.Rev.}
  {\bfseries D85} (2012) 126002},
\href{http://arxiv.org/abs/1203.0755}{{\ttfamily arXiv:1203.0755 [hep-th]}}.
%%CITATION = ARXIV:1203.0755;%%.

\bibitem{1302.0697}
M.~P. Heller, R.~A. Janik, and P.~Witaszczyk, ``{Hydrodynamic Gradient
  Expansion in Gauge Theory Plasmas},''
  \href{http://dx.doi.org/10.1103/PhysRevLett.110.211602}{{\em Phys.Rev.Lett.}
  {\bfseries 110} (2013) 211602},
\href{http://arxiv.org/abs/1302.0697}{{\ttfamily arXiv:1302.0697 [hep-th]}}.
%%CITATION = ARXIV:1302.0697;%%.

\bibitem{1411.1969}
J.~Jankowski, G.~Plewa, and M.~Spalinski, ``{Statistics of thermalization in
  Bjorken Flow},'' \href{http://dx.doi.org/10.1007/JHEP12(2014)105}{{\em JHEP}
  {\bfseries 1412} (2014) 105},
\href{http://arxiv.org/abs/1411.1969}{{\ttfamily arXiv:1411.1969 [hep-th]}}.
%%CITATION = ARXIV:1411.1969;%%.

\bibitem{0712.2451}
R.~Baier, P.~Romatschke, D.~T. Son, A.~O. Starinets, and M.~A. Stephanov,
  ``{Relativistic viscous hydrodynamics, conformal invariance, and
  holography},'' \href{http://dx.doi.org/10.1088/1126-6708/2008/04/100}{{\em
  JHEP} {\bfseries 0804} (2008) 100},
\href{http://arxiv.org/abs/0712.2451}{{\ttfamily arXiv:0712.2451 [hep-th]}}.
%%CITATION = ARXIV:0712.2451;%%.


\bibitem{data} \url{http://physics.bgu.ac.il/~lublinm/Data.nb}


\bibitem{1409.5087}
M.~P. Heller, R.~A. Janik, M.~Spalinski, and P.~Witaszczyk, ``{Coupling
  hydrodynamics to nonequilibrium degrees of freedom in strongly interacting
  quark-gluon plasma},''
  \href{http://dx.doi.org/10.1103/PhysRevLett.113.261601}{{\em Phys.Rev.Lett.}
  {\bfseries 113} (2014) 261601},
\href{http://arxiv.org/abs/1409.5087}{{\ttfamily arXiv:1409.5087 [hep-th]}}.
%%CITATION = ARXIV:1409.5087;%%.

\bibitem{hep-th/0002230}
S.~de~Haro, S.~N. Solodukhin, and K.~Skenderis, ``{Holographic reconstruction
  of space-time and renormalization in the AdS / CFT correspondence},''
  \href{http://dx.doi.org/10.1007/s002200100381}{{\em Commun.Math.Phys.}
  {\bfseries 217} (2001) 595--622},
\href{http://arxiv.org/abs/hep-th/0002230}{{\ttfamily arXiv:hep-th/0002230
  [hep-th]}}.
%%CITATION = HEP-TH/0002230;%%.

\bibitem{hep-th/0010138}
K.~Skenderis, ``{Asymptotically Anti-de Sitter space-times and their stress
  energy tensor},'' \href{http://dx.doi.org/10.1142/S0217751X0100386X}{{\em
  Int.J.Mod.Phys.} {\bfseries A16} (2001) 740--749},
\href{http://arxiv.org/abs/hep-th/0010138}{{\ttfamily arXiv:hep-th/0010138
  [hep-th]}}.
%%CITATION = HEP-TH/0010138;%%.

\bibitem{hep-th/9902121}
V.~Balasubramanian and P.~Kraus, ``{A Stress tensor for Anti-de Sitter
  gravity},'' \href{http://dx.doi.org/10.1007/s002200050764}{{\em
  Commun.Math.Phys.} {\bfseries 208} (1999) 413--428},
\href{http://arxiv.org/abs/hep-th/9902121}{{\ttfamily arXiv:hep-th/9902121
  [hep-th]}}.
%%CITATION = HEP-TH/9902121;%%.

\bibitem{1304.2433}
M.~Hindmarsh, S.~J. Huber, K.~Rummukainen, and D.~J. Weir, ``{Gravitational
  waves from the sound of a first order phase transition},''
  \href{http://dx.doi.org/10.1103/PhysRevLett.112.041301}{{\em Phys.Rev.Lett.}
  {\bfseries 112} (2014) 041301},
\href{http://arxiv.org/abs/1304.2433}{{\ttfamily arXiv:1304.2433 [hep-ph]}}.
%%CITATION = ARXIV:1304.2433;%%.

\bibitem{1412.5147}
T.~Kalaydzhyan and E.~Shuryak, ``{Gravity waves generated by sounds from Big
  Bang phase transitions},''
\href{http://arxiv.org/abs/1412.5147}{{\ttfamily arXiv:1412.5147 [hep-ph]}}.
%%CITATION = ARXIV:1412.5147;%%.

\bibitem{hep-th/9806087}
M.~Henningson and K.~Skenderis, ``{The Holographic Weyl anomaly},''
  \href{http://dx.doi.org/10.1088/1126-6708/1998/07/023}{{\em JHEP} {\bfseries
  9807} (1998) 023},
\href{http://arxiv.org/abs/hep-th/9806087}{{\ttfamily arXiv:hep-th/9806087
  [hep-th]}}.
%%CITATION = HEP-TH/9806087;%%.

\bibitem{0801.3701}
R.~Loganayagam, ``{Entropy Current in Conformal Hydrodynamics},''
  \href{http://dx.doi.org/10.1088/1126-6708/2008/05/087}{{\em JHEP} {\bfseries
  0805} (2008) 087},
\href{http://arxiv.org/abs/0801.3701}{{\ttfamily arXiv:0801.3701 [hep-th]}}.
%%CITATION = ARXIV:0801.3701;%%.

\bibitem{0704.1647}
M.~Lublinsky and E.~Shuryak, ``{How much entropy is produced in strongly
  coupled Quark-Gluon Plasma (sQGP) by dissipative effects?},''
  \href{http://dx.doi.org/10.1103/PhysRevC.76.021901}{{\em Phys.Rev.}
  {\bfseries C76} (2007) 021901},
\href{http://arxiv.org/abs/0704.1647}{{\ttfamily arXiv:0704.1647 [hep-ph]}}.
%%CITATION = ARXIV:0704.1647;%%.

\bibitem{hep-th/0602059}
P.~Kovtun and A.~Starinets, ``{Thermal spectral functions of strongly coupled
  N=4 supersymmetric Yang-Mills theory},''
  \href{http://dx.doi.org/10.1103/PhysRevLett.96.131601}{{\em Phys.Rev.Lett.}
  {\bfseries 96} (2006) 131601},
\href{http://arxiv.org/abs/hep-th/0602059}{{\ttfamily arXiv:hep-th/0602059
  [hep-th]}}.
%%CITATION = HEP-TH/0602059;%%.

\bibitem{hep-th/0205051}
D.~T. Son and A.~O. Starinets, ``{Minkowski space correlators in AdS / CFT
  correspondence: Recipe and applications},''
  \href{http://dx.doi.org/10.1088/1126-6708/2002/09/042}{{\em JHEP} {\bfseries
  0209} (2002) 042},
\href{http://arxiv.org/abs/hep-th/0205051}{{\ttfamily arXiv:hep-th/0205051
  [hep-th]}}.
%%CITATION = HEP-TH/0205051;%%.

\bibitem{BLS} Y. Bu, M. Lublinsky, and A. Sharon, in preparation



\end{thebibliography}
\end{document}